\documentclass[onecolumn,authoryear]{els-mrw} 

\usepackage{amsmath,amssymb,amsfonts,amsthm,makeidx,graphicx}
\usepackage{txfonts,array,multirow}
\usepackage{helvet}

\newcommand{\bl}[1]{{\color{blue} #1}}
\newcommand{\rb}[1]{\raisebox{1.5ex}[-1.5ex]{#1}}
\newcommand{\rbb}[1]{\raisebox{2.5ex}[-1.5ex]{#1}}

\begin{document}

\chapter{Hadron Spectroscopy: Experimental Overview
\label{chap:HadronSpectroscopy}}

\author[1]{Volker Crede}%


\address[1]{\orgname{\,Florida State University}, \orgdiv{Department of Physics},
  \orgaddress{77 Chieftan Way, Tallahassee, FL 32306~~USA}}


\maketitle

\begin{glossary}[Key Points]
  \term{The Standard Model of Particle Physics} is the theory that describes all fundamental
  particles and their interactions (except gravity).\\[0.4ex]
  
  \term{Color Confinement} refers to the observation that no free quarks have
    been observed. They are rather confined in hadrons.\\[0.4ex]
  
  \term{Unconventional Hadrons} are strongly interacting particles that do not fit neatly
  into the traditional quark-model picture of hadrons.\\[0.4ex]
  
\end{glossary}


\begin{abstract}[Abstract]
One of the most fascinating phenomena in nuclear and particle physics is the formation of light
hadrons -- strongly interacting particles -- out of massless gluons and almost massless quarks.
The proton and the neutron are the most prominent examples of a hadron. These particles exhibit
rich excitation spectra due to their complex quark-gluon structures. Short-lived pairs of virtual
quarks and antiquarks are continually formed and annihilated inside hadrons. Every physics student
probably knows that the proton consists of three quarks. This statement is misleading, though. For
example, the proton only shows a constant excess of three quarks versus antiquarks from the outside,
and these three quarks are not even well defined. Understanding hadrons goes well beyond explaining
their properties in terms of three quarks for baryons and a quark-antiquark pair for mesons. Much like
atomic spectroscopy revealed the structure of atoms through discrete energy levels, hadron spectroscopy
seeks the mapping of the spectrum of hadronic states and to use that information to better understand
the underlying theory of the strong force -- Quantum Chromodynamics (QCD). Over the past few decades,
hadron spectroscopy has entered an exciting new era with the discovery of many exotic hadrons that do
not fit neatly into the traditional quark model classification of ordinary baryons and mesons. Many of the
discoveries come from the heavy-flavor sector. This renaissance in hadron spectroscopy has challenged
our understanding of hadrons but also opened new windows of opportunity in hadron research. Despite
significant progress, many questions remain unanswered, particularly regarding the nature of exotic
hadrons and the full implications of QCD in the non-perturbative regime. As experimental techniques
and computational methods continue to advance, hadron spectroscopy will remain a vital tool for probing
the deepest layers of matter and understanding the fundamental structure of the universe.
\end{abstract}

\begin{figure}[b]
  \centering
  \begin{minipage}[b]{0.35\textwidth}
      \centering
      \includegraphics[width=0.72\textwidth]{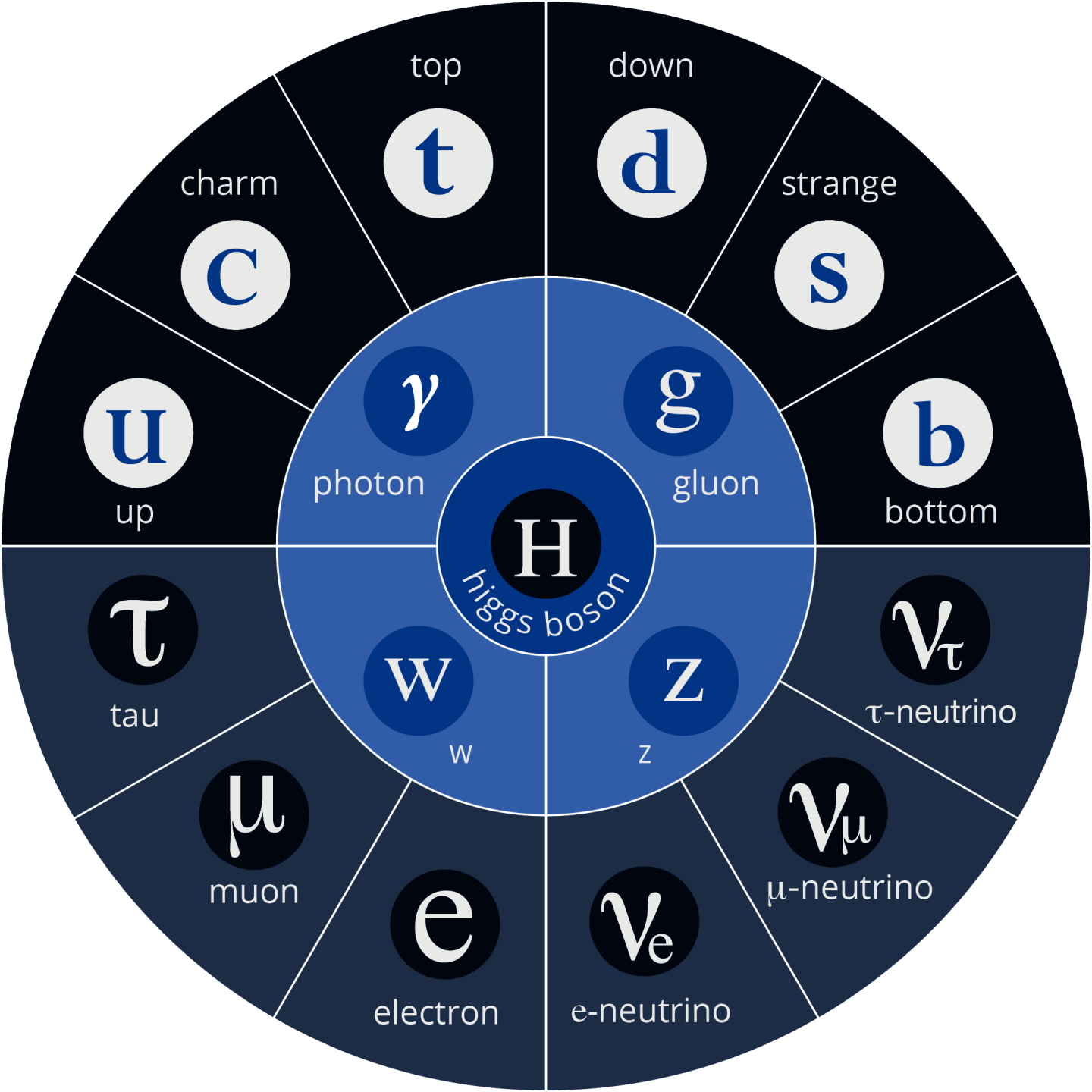}
  \end{minipage}
  \hfill
  \begin{minipage}[t]{0.63\textwidth}\vspace{-3.4cm}
    \centering
    {\addtolength{\extrarowheight}{4pt}
      \begin{tabular}{|c|ccc|c|c|c|c|}\hline
        & \multicolumn{3}{c|}{Generations} & & & Baryon & Lepton\\
        & 1 & 2 & 3 & \rb{Electric Charge} & \rb{Spin} & Number & Number\\[0.5ex]\hline
        & $\nu_{\rm e}$ & $\nu_\mu$ & $\nu_\tau$ & 0 & & &\\
        \rb{Leptons} & e & $\mu$ & $\tau$ & $-1$ & \rb{1/2} & \rb{0} & \rb{1}\\[0.5ex]\hline
         & u & c & t & $+2/3$ & & &\\
        \rb{Quarks} & d & s & b & $-1/3$ & \rb{1/2} & \rb{1/3} & \rb{0}\\[0.5ex]\hline
      \end{tabular}}
      \vspace{0.3cm}
  \end{minipage}
\caption{The fundamental particles of the Standard Model and the basic properties of the elementary
  fermions. Three generations of quarks and leptons exist. In addition, all ordinary SM fermions come
  with a corresponding antiparticle that has identical mass and spin. However, antiparticles possess
  opposite electric charge as well as opposite baryon and lepton numbers.\label{chap1:fig1}}
\end{figure}

\section{Introduction\label{HadronSpectroscopy:sec1}}
Perhaps one of the greatest accomplishments in modern physics has been the so-called Standard
Model (SM) of particle physics which describes three of the four known fundamental interactions in
the universe (weak, strong and electromagnetic, excluding gravitational forces) and classifies all
known elementary particles. In the SM, twelve fermions, six quarks and six leptons, form three generations
of matter particles. In addition, four vector bosons ($W$, $Z$~bosons, photon, gluon) mediate the interactions
among the elementary particles, and the Higgs boson helps explain why all the other particles have mass.
The classification scheme of the elementary particles in the SM is shown in Fig.~\ref{chap1:fig1}. Six types
of quarks, known as flavors and named {\it up}, {\it down}, {\it strange}, {\it charm}, {\it bottom}, and
{\it top}, are experimentally established, with the discovery of the top quark in 1995 at Fermilab completing
the picture, see for example Ref.~[\cite{D0:1995jca}]. The quarks are the fundamental building blocks of all
visible matter. They combine to form composite particles called hadrons, and the gluons serve as the
mediators of the strong nuclear forces {\it binding} them together. Both quarks and gluons carry the strong
color charges, commonly known as {\it red}, {\it green}, and {\it blue}. However, unlike electric charge, color
charge is never observed in nature. Rather, color charges combine to form {\it color-neutral} (or {\it white}) objects,
i.e., all three colors can combine to form white, or any color and its anti-color. It is worth noting that the term
``color charge'' is entirely unrelated to the human perception of color, which refers to the frequency of
electromagnetic radiation. The nuclear color charge in terms of red, green, and blue merely provides a
convenient analogy to the regular additive primary colors.

Quantum chromodynamics (QCD) provides the theoretical framework of the strong nuclear force.
Developed in the 1970s based on the concept of color as the source of a {\it strong field,} this quantum
field theory is well proven and allows physicists to make very precise predictions of high-energy physics
phenomena at very short distance scales.
\begin{BoxTypeA}[chap1:box1]{QCD exhibits three distinct features:}
\begin{enumerate}
  \item {\bf Asymptotic freedom} as a property of QCD was discovered in the 1970s and describes the
    weakening interaction strength between quarks toward increasing momentum transfers or short
    distance scales. The 2004 Nobel Prize in Physics was dedicated to the study of asymptotic freedom;
    see Ref.~[\cite{Gross:1973id}] for the original discovery paper. For large exchanges of momentum
    (a scenario that probes the short-distance behavior of the quarks, in simple terms, because of the
    inverse relationship between the exchanged momentum and the {\it de Broglie} wavelength), QCD is
    amenable to perturbation theory as a mathematical tool to find approximate solutions from first principles.
  \item {\bf Chiral symmetry breaking} plays a central role in low-energy hadron physics. The term refers
    to the chiral symmetry of the QCD Lagrangian, which describes the independence of left- and
    right-handed quark fields for massless quarks under chiral transformations. This symmetry is
    spontaneously broken and accompanied by a qualitative rearrangement of the vacuum ground state
    characterized by the formation of a non-zero quark condensate, i.e., the vacuum is now populated
    with scalar quark-antiquark pairs: The vacuum does not respect the symmetry of the underlying theory.
    The chiral symmetry is also explicitly broken because the quarks have non-zero masses. As a consequence
    of chiral symmetry breaking, light pseudoscalar mesons emerge as approximate Nambu–Goldstone bosons.
    Quarks can also develop large effective masses when exposed to the condensate (dynamical symmetry
    breaking), which may explain the origin of their effective or constituent masses.
  \item {\bf Color confinement} describes the ``dogma'' in particle physics that individual quarks
    are never observed in nature. In other words, color-charged particles cannot be isolated, and a quark
    cannot be removed from a hadron. The phenomenon can be qualitatively understood by noting that
    the force-carrying gluons of QCD have color charge, unlike the photons of quantum electrodynamics
    (QED) that do not carry a positive or negative electric charge. The electric field between electrically
    charged particles decreases rapidly as those particles are separated, whereas the gluon field between
    a pair of color charges does not weaken with an increasing separation; rather, the confining force between
    the quarks is approximately constant at large separations, leading to a potential energy that increases
    linearly with distance. Although analytically unproven, color confinement is well established from lattice-QCD
    calculations and decades of experiments. This is the realm of non-perturbative or strong-coupling QCD.
    The structure of hadrons has a non-perturbative nature, rendering standard perturbation theory inapplicable
    to describe hadron properties.
\end{enumerate}
\end{BoxTypeA}  
Hadrons can be broadly grouped into two primary categories: mesons and baryons. Mesons are bosons and
consist of an even number of ``excess'' or valence
quarks that define the overall flavor content, one quark and one
antiquark in the simplest case for conventional mesons. Baryons are
fermions and contain an odd number of valence quarks, three quarks for
regular baryons. The proton with a $(uud\,)$~valence quark structure is
the lightest and certainly most prominent example of a
baryon. However, QCD also allows for other color-neutral systems such as
multiples of $(q\bar{q})$. The lightest such system would be a
$(q\bar{q}q\bar{q})$ tetraquark. In addition, mesons with valence glue are
possible $(q\bar{q}g)$, so-called hybrid mesons, or objects that consist entirely of gluons without valence
quarks (glueballs). Such unconventional objects (tetraquarks, hybrid
mesons, glueballs, etc.) are still considered mesons since they carry {\it zero} baryon
number $(B=0)$, see Fig.~\ref{chap1:fig1} (right side). However, they can have spin-parity
quantum numbers that are not possible for
conventional $(q\bar{q})$~mesons. Baryons can also contain valence glue
$(qqqg)$, but the lack of exotic quantum numbers that are not possible
for conventional baryons makes them hard to identify in
nature. Pentaquarks consisting of $(qqq\bar{q}q)$ or the possibility of a
baryon-meson molecule $(qqq)(q\bar{q})$ are intriguing and have become talking points
since the discovery of the 2017 pentaquark candidates, $P_c(4380)^+$
and $P_c(4450)^+$, in $\Lambda_b^0\to J/\psi p K^-$~decays by the LHCb Collaboration at
CERN~[\cite{LHCb:2015yax}]. 

All hadrons can exist in various excited states, each with distinct masses, lifetimes, and quantum numbers.
Hadron spectroscopy aims to identify these states, measure their
properties, and organize them into a coherent classification
scheme. A major experimental and theoretical challenge 
is the highly non-linear and strongly coupled nature of QCD at low energies. For his reason,
standard perturbative techniques, which work well in other areas
of quantum field theory, are not applicable. Instead, physicists rely
on a combination of experimental data, phenomenological models, and
numerical simulations such as lattice QCD to study hadronic
systems. Lattice QCD, in particular, involves discretizing spacetime
into a grid and performing large-scale numerical computations to
approximate the behavior of quarks and gluons from first
principles. We refer to the literature and other chapters of the Encyclopedia of Nuclear
Physics for a theoretical overview of
hadron spectroscopy. 

\section{Spectroscopy of Meson Systems: An Experimental Overview
  \label{MesonSpectroscopy:sec2}}
Mesons are composed of quarks,
antiquarks, and gluons such that the net baryon number of the
state is zero ($B = 0$). One of the earliest and most influential tools in hadron spectroscopy
is the quark model, developed in the 1960s. This model classifies
hadrons based on their quark content and quantum numbers such as spin,
parity, and isospin. It successfully organized a large number of known
particles into families and also predicted the existence of new ones, many
of which were later experimentally confirmed. Despite its simplicity,
the quark model remains a useful guide for the community, although it is understood
to be an approximation of the more complex quark-gluon dynamics in QCD.
In the traditional quark model, conventional mesons are described as $q\bar{q}$ bound
states of a quark and an antiquark, and can be considered the nuclear analogue of 
positronium $(e^+e^-)$. Mesons containing only $u$, $d$, and $s$ quarks are referred to as
light-flavor mesons, whereas mesons containing $c$ and $b$ quarks are known as
heavy-flavor mesons. Spectroscopy of mesons containing $t$ quarks
is merely of theoretical nature since the $t$~quark's lifetime is shorter than the
typical hadronization time scale. As a result, the $t$~quark decays
via the weak interaction before it can bind with other quarks to form hadrons.

The dominant reference source for all known particles is the Review of
Particle Physics (RPP) by the Particle Data Group (PDG) which is published on a
biennial basis and updated annually online~[\cite{ParticleDataGroup:2024cfk}]. This international
collaboration maintains the listings of all matter particles including
leptons, quarks, mesons and baryons, summarizes the latest experimental results
on particle properties, and reviews the current status of the Standard
Model physics. First published in 1964 as {\it Data on Elementary
  Particles and Resonant States}, the RPP also publishes summaries of
fundamental mathematical and statistical concepts, and regularly
updated mini-reviews on various topics relevant for hadron spectroscopy, including
the nature of individual resonances.

The following sections will give a brief introduction to the quantum numbers and
the naming scheme of mesons. Furthermore, experimental methods will be
presented for the study of conventional and exotic meson(-like)
systems. The subsequent discussion of known mesons is not considered a
comprehensive review of the field for the expert but will rather provide an
overview of an evolving field, driven by rapid advancements in
computing and experimental techniques, including hardware developments
and AI-driven analysis tools.

\subsection{Meson quantum numbers\label{MesonQN:subsec1}}
In the 1960s, Gell-Mann and Ne'eman independently introduced the
so-called eightfold way as an organizational scheme for the overwhelming
number of particles, including hadrons, that had been observed at the
time (``particle zoo''). Based on the representation theory of SU(3), the proposed scheme
quickly evolved into flavor-SU(3) and the development of the quark
model. The strong nuclear force is flavor-independent and therefore,
flavor symmetry approximately holds for the three light $u$, $d$, and $s$~quarks since their masses
are similar and smaller than the strong interaction scale.

\begin{figure}[t]
  \centering
  \begin{minipage}[b]{0.42\textwidth}
      \centering
      \includegraphics[width=1.025\textwidth]{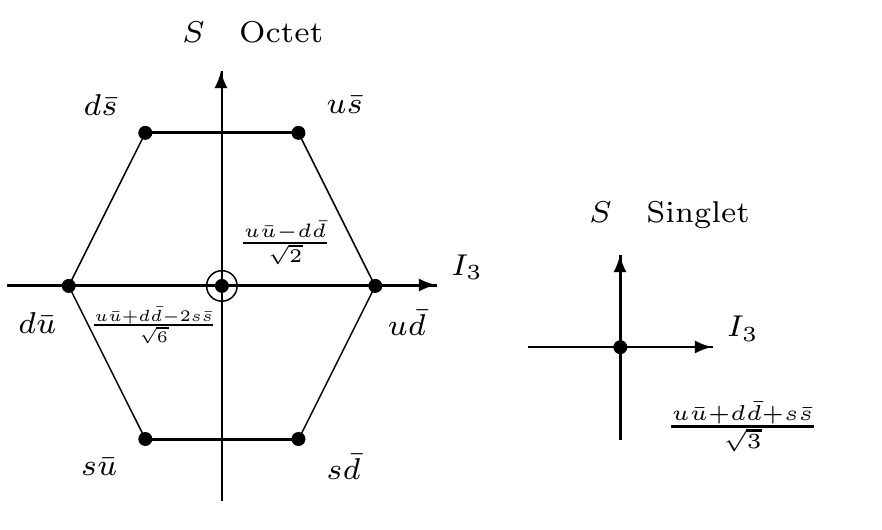}
  \end{minipage}
  \hspace{0.5cm}
  \begin{minipage}[b]{0.42\textwidth}
      \centering
      \includegraphics[width=0.656\textwidth]{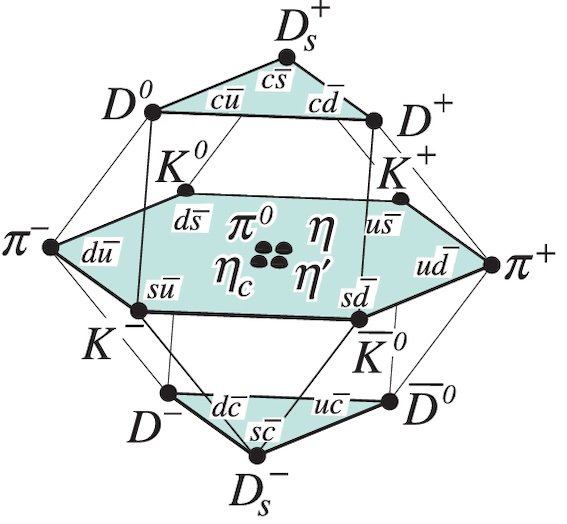}
  \end{minipage}
  \caption{SU(3)$_f$ and SU(4)$_f$ multiplets. {\bf Left side:} The light
    $u$, $d$, and $s$ quarks and their corresponding antiquarks,
    $\bar{u}$, $\bar{d}$, $\bar{s}$, form the basis for ${\bf
  3}\otimes{\bf \bar{3}} = {\bf 8}\oplus {\bf 1}$ mesons. {\bf Right
  side:} SU(4)$_f$ weight diagram of the pseudoscalar mesons in the ${\bf 4}\otimes{\bf \bar{4}} = {\bf
  15}\oplus {\bf 1}$ representation. In addition
to isospin $I_3$ and strangeness $S$, the diagram is a function of
charm $C$ in the vertical.\label{chap2:fig1}}
\end{figure}

These three light quarks and the corresponding three light antiquarks can form nine possible
$q\bar{q}$ combinations. Following the rules of the representation theory SU(3), these nine
combinations can decompose into an octet and a singlet: ${\bf
  3}\otimes{\bf \bar{3}} = {\bf 8}\oplus {\bf 1}$. These multiplets
are shown in Fig.~\ref{chap2:fig1} (left side) in the
usual representation of {\it strangeness, S,}\,\footnote{The strange
  quark has {\it strangeness} $S=-1$, the charm quark has {\it charm}
  $C=+1$, the bottom quark has {\it bottomness} $B=-1$, and the top
  quark has {\it topness} $T=+1$. The flavor sign is the same as the
  sign of the electric charge.}
versus the third component of {\it isospin, $I_3$}. The latter quantum
number is related to the $u$- and $d$-quark content of the hadron,
$I_3 = 1/2\,(n_u - n_d)$,
where isospin is defined as a vector quantity with $I = 1/2$, and $I_3 =
+1/2$ as well as $I_3 = -1/2$ for the $u$
quark and $d$ quark, respectively. The isospin for all other quarks is
$I=0$. From a practical point of view, the isospin defines the number
of a hadron's charged states as the number of projections of its
isospin vector. For example, the $\pi$~meson has $I=1$ and comes in
$2I+1$ charged states, that is, as an isospin triplet with charged states $\pi^+$, $\pi^-$,
$\pi^0$. Shown in Fig.~\ref{chap2:fig1} is also the flavor content of the multiplet
members. A fourth $c$ or $b$
quark can be included by extending flavor-SU(3) to flavor-SU(4), but
the latter symmetry is badly broken due to the much heavier mass
of the additional quark compared to the lighter $u$, $d$, and $s$
quarks. The 16 possible combinations decompose into
${\bf 4}\otimes{\bf \bar{4}} = {\bf 15}\oplus {\bf 1} =  {\bf
  8}\oplus {\bf 3}\oplus {\bf \bar{3}}\oplus {\bf 1}\oplus {\bf 1}$, and these multiplets contain the ${\bf
  8}\oplus {\bf 1}$ nonet of light flavors.
Figure~\ref{chap2:fig1} (right side) shows the SU(4)$_f$~multiplets
and, as an example, the nominal mapping of these states onto the
familiar pseudoscalar mesons.
The SU(3)$_f$~nonet appears as the middle layer of the {\bf 16}-plet.

Historically, mesons were classified based on the traditional
non-relativistic quark model. Within this framework, the spectroscopic
state of the $q\bar{q}$~system was described in terms of $L$ and $S$,
the relative orbital angular momentum between the quarks and the intrinsic
spin~$S$, respectively. The total intrinsic spin of the
$q\bar{q}$~system can either be $S = 0$ or $S = 1$, which results
in the total angular momentum of $\vec{J} = \vec{L} \oplus
\vec{S}$, where $L$ and $S$ quantum-mechanically add as two
angular momenta. Orbital angular momentum $L$ is considered an independent
approximate quantum number for mesons when spin-orbit interactions are
small or in the description of the ground
state or lower-lying excited states. However, in strongly interacting environments, spin-orbit
interactions can couple $L$ and $S$, causing them not to be conserved independently. The total
angular momentum $J$ remains a good quantum number, though. Moreover,
relativistic effects can lead to observer-dependent values of $L$, and in excited states, different 
$L$~states can mix, and thus, the assignment of a single 
$L$~value is less accurate.

\begin{BoxTypeA}[subsec1:box1]{Quantum numbers of light-flavor mesons:}
\begin{enumerate}   
  \item {The intrinsic \bf{parity, $P$,}} must be
     negative for a quark-antiquark system since the parity of the antiquark is opposite
     to that of the quark. For this reason, the total parity of the $q\bar{q}$~system is
$P_{q\bar{q}} = -(-1)^L$, where $L$ is the relative angular momentum
between the two quarks.
  \item {\bf{Charge conjugation, $C$,}} results from a discrete symmetry
     transformation that replaces all particles with their
     corresponding antiparticles. Thus, for a $q\bar{q}$~system, only
     electrically neutral states can be eigenstates of the
     $\hat{C}$~operator and without discussing any details, the $C$-parity is $C_{q\bar{q}} = (-1)^{L+S}$.
   \item {\bf{$\boldsymbol{G}$ parity}} is an additional quantum number for 
     any $q\bar{q}$~system independent of charge and can be defined as an extension of $C$~parity. Consider an
    isovector state with $I = 1$, e.g., the pion. The
    $\hat{C}$~operator would transform a $\pi^\pm$ to its
    corresponding state with opposite electric charge. A rotation in
    isospin space can then transform this back to its original
    state. The $G$~parity is therefore defined as $G = C\,(-1)^I =
    (-1)^{L+S+I}$.
\end{enumerate}
\end{BoxTypeA}

\begin{table}[b]
  \centering
  \small
      \caption{Tentative assignments for some of the observed conventional mesons ($L\leq 2$)
        as suggested by the PDG~[\cite{ParticleDataGroup:2024cfk}]. States labeled
        $^\dagger$ are not considered established.
      Mesons labeled $^a$ split into two states in some models,
      and mesons labeled $^b$ have also been interpreted as
      tetraquarks or $D^{(\ast)} K$~molecules. Furthermore,
      $D_{s1}^\ast(2700)^\pm$ is a mixture of both the $1\,^3D_1$ and
      $2\,^3S_1$~states with $J^{PC} = 1^{--}$. Finally, an
      interpretation of $\psi(4160)$ as a single resonance is
      questionable due to some threshold effects in this energy region.\label{chap2:tab1}}
      {\addtolength{\extrarowheight}{4pt}
        \begin{tabular}{@{}cc|c|cc|c||cc|cc|cc|c@{}}
        \multicolumn{12}{c}{}\\[-3ex]\hline
        & & & & & & \multicolumn{2}{c|}{$I=0$}
        & $I=1/2$ & $I=0$ & $I=1/2$ & $I=0$ & $I=0$\\
        \rb{$~~n\,^{2S+1}L_J$} & \rb{$J^{PC}$} & \rb{$I=1$} &
                                                       \multicolumn{2}{c|}{\rb{$I=0$}} & \rb{$I=\frac{1}{2}$} & $c\bar{c}$ & $b\bar{b}$
        & $c\bar{q},\,\bar{c}q$ & $c\bar{s};\,\bar{c}s$ &
                                                          $b\bar{q},\,\bar{b}q$ & $b\bar{s};\,\bar{b}s$ & $b\bar{c};\,\bar{b}c$\\\hline
        $1\,^1S_0$ & $0^{-+}$ & $\pi$ & $\eta$ & $\eta^{\,\prime}$ & $K$
        & $\eta_c(1S)$ & $\eta_b(1S)$ & $D$ & $D^{\,\pm}_s$ & $B$ & $B^{\,0}_s$ & $B_c^{\,\pm}$\\ 
        $1\,^3S_1$ & $1^{--}$ & $\rho$ & $\omega$ & $\phi$ & $K^{\,\ast}$ & $J/\psi$  
        & $\Upsilon(1S)$ & $D^{\,\ast}$ & $D^{\,\ast\,\pm}_s$ & $B^{\,\ast}$ & $B^{\,\ast}_s$ &\\
        $1\,^1P_1$ & $1^{+-}$ & $b_1$ & $h_1$ & $h^{\,\prime}_1$ & $K_{1B}$ & $h_c(1P)$
        & $h_b(1P)$ & $D_1(2420)$ & $D_{s1}(2536)^\pm$ & $B_1(5721)$ & $B_{s1}(5830)^0$ &\\
        $1\,^3P_0$ & $0^{++}$ & $a_0$ & $f_0$ & $f_0^{\,\prime}$ & $K_0^{\,\ast}$
        & $\chi_{c0}(1P)$ & $\chi_{b0}(1P)$ & $D_0^{\,\ast}(2300)^{\,a}$ & $D^{\,\ast}_{s0}(2317)^{\pm\,b}$ & & &\\ 
        $1\,^3P_1$ & $1^{++}$ & $a_1$ & $f_1$ & $f_1^{\,\prime}$ & $K_{1A}$ & $\chi_{c1}(1P)$  
        & $\chi_{b1}(1P)$ & $D_1(2430)^{0\,a}$ & $D_{s1}(2460)^{\pm\,b}$ & & &\\
        $1\,^3P_2$ & $2^{++}$ & $a_2$ & $f_2$ & $f_2^{\,\prime}$ & $K_2^\ast$ & $\chi_{c2}(1P)$
        & $\chi_{b2}(1P)$ & $D_2^{\,\ast}(2460)$ & $D_{s2}^{\,\ast}(2573)$ & $B_2^{\,\ast}(5747)$ &
        $B_{s2}^{\,\ast}(5840)^0$ &\\
        $1\,^1D_2$ & $2^{-+}$ & $\pi_2$ & $\eta_2$ & $\eta_2^{\,\prime}$ & $K_2$ & & & & & & &\\ 
        $1\,^3D_1$ & $1^{--}$ & $\rho_1$ & $\omega_1$ & $\phi_1$ & $K_1^\ast$ & $\psi(3770)$  
        & & $D_1^{\,\ast}(2760)^{0\,\dagger}$ & $D_{s1}^{\,\ast}(2860)^{\pm\,\dagger}$ & & &\\
        $1\,^3D_2$ & $2^{--}$ & $\rho_2$ & $\omega_2$ & $\phi_2$ & $K_2$ & $\psi_2(3823)$
        & $\Upsilon_2(1D)$ & $D_2(2740)^{0\,\dagger}$ & & & &\\
        $1\,^3D_3$ & $3^{--}$ & $\rho_3$ & $\omega_3$ & $\phi_3$ & $K_3^\ast$ & $\psi_3(3842)$
        & & $D_3^{\,\ast}(2750)$ & $D_{s3}^{\,\ast}(2860)$ & & &\\[0.6ex]\hline
        $2\,^1S_0$ & $0^{-+}$ & $\pi$ & $\eta$ & $\eta$ & $K$ & $\eta_c(2S)$
        & $\eta_b(2S)^\dagger$ & $D_0(2550)^{0\,\dagger}$ & $D_{s0}(2590)^{+\,\dagger}$ & & & $B_c(2S)^\pm$\\
        $2\,^3S_1$ & $1^{--}$ & $\rho$ & $\omega$ & $\phi$ & $K^\ast$ & $\psi(2S)$
        & $\Upsilon(2S)$ & $D_1^\ast(2600)^{0\,\dagger}$ & $D_{s1}^\ast(2700)^\pm$ & & &\\
        $2\,^1P_1$ & $1^{+-}$ & & & & & & $h_b(2P)$ & & & & &\\
        $2\,^3P_J$ & $J^{\,++}$ & & & & & $\chi_{cJ}(2P)$ & $\chi_{bJ}(2P)$ & $D_2^\ast(3000)^{0\,\dagger}$
                          & & & &\\
        $2\,^3D_1$ & $1^{--}$ & & & & & $\psi(4160)^{\rm }$ & & & & & &\\
        $3\,^3S_1$ & $1^{--}$ & & & & & $\psi(4040)$ & $\Upsilon(3S)$ & & & & &\\
        $4\,^3S_1$ & $1^{--}$ & & & & & $\psi(4415)$ & $\Upsilon(4S)$ & & & & &\\[0.6ex]\hline
    \end{tabular}}
\end{table}

\subsection{Naming scheme of mesons}
Prior to 1986, no systematic naming scheme for
mesons existed. Those who discovered new states often proposed
what those states would be called, and many Greek symbols were used.
As an example, the first meson to be discovered in 1947 was the
$\pi$~meson. Legend has it that Yukawa chose the Greek letter $\pi$
because of its resemblance to a Chinese character that means ``mediate,''
referring to the $\pi$~meson's role as the strong force mediator
between nucleons in the atomic nucleus.  
In addition to maintaining particle listings and reviews, the Particle
Data Group follows a prescriptive philosophy and proposes standards
and best practices. Therefore, the PDG introduced a naming scheme for
mesons in 1986 that is
still in use today. This scheme assigns light mesons to nonets with definite $J^{PC}$~quantum
numbers in addition to the isospin of the nonet
member. Thus, the base name is the same for
all mesons of a given $I$ and $PC$ combination, with an additional subscript
denoting the total spin $J$. Table~\ref{chap2:tab1} shows the naming
scheme for light- and heavy-flavor mesons (including observed states).

\subsection{Experimental methods for studying (light) meson and
  meson-like systems}
Studying meson resonances across various production mechanisms provides
crucial clues about the nature of observed states. Conservation of
certain quantities, couplings or cross sections, and available energies impose
constraints on the production process. All major experimental
methods in meson spectroscopy can broadly be grouped into three broad
categories.
\subparagraph{1. Strong production}
Mesons are copiously produced in scattering experiments based on the
strong nuclear interaction. Hadronic cross sections are large compared
to electromagnetic cross sections, but
these typically come at the expense of larger backgrounds. Examples include high-energy scattering
experiments in which meson or proton beams are incident on nucleon or
nuclear targets, leading to diffractive dissociation under small
momentum exchanges or charge exchange processes,
as well as central production reactions in hadron-hadron scattering. This group of
strong-production experiments also includes proton-antiproton annihilation, where the
momentum of the antiproton beam can range from close to 0~MeV
(proton-antiproton annihilation at rest) to fairly high momenta of
several GeV. The major players of hadron-beam experiments in 2026 are
COMPASS at CERN and the experiments at the Large Hadron Collider (LHC)
at CERN, LHCb in particular.

\subparagraph{2. Electromagnetic production}
Complementary to meson production induced by strongly interacting
hadrons, electromagnetic probes in the form of photon or lepton beams can
be used to study meson resonances. For example, electromagnetic production can be
studied in fixed-target experiments, e.g., in $\gamma\,(e^-)$ + proton
reactions (so-called photo- or electroproduction), but also in collider
experiments using $e^+e^-$ scattering reactions ranging from annihilation to
two-photon fusion processes, and meson production in initial-state radiation.
In 2026, the major active experiments at $e^+e^-$~collider facilities are BES\,III at the Beijing
Electron–Positron Collider~II (BEPC~II) in China operating in the
$\tau$-charm energy range, Belle~II at
SuperKEKB in Japan collecting data largely on the
$\Upsilon(4S)$~resonance, and SND \& CMD-3 at VEPP-2000 in Russia
operating at energies below 2~GeV. Photoproduction experiments are
performed by the CLAS12 and GlueX Collaborations at Jefferson
Laboratory in the United States, where the high-quality electron beam
provided by the Continuous Electron Beam Accelerator Facility (CEBAF) enables precision
investigations of meson production from the nucleon. 

\subparagraph{3. Heavy-flavor decays into (light) mesons}
Heavy-flavor resonances are typically produced at the high-energy
facilities because they require higher-energy beams. Here, the decays of $J/\psi$, $\psi(2S)$, $D$,
$D_s$, or even $\eta_c$ into light mesons provide a great laboratory. 
In a similar way, transitions from vector charmonium-like states
provide access to heavy mesons, to $\chi_{c1}(3872)$ and some of the
new alleged tetraquark states, for instance. Moreover, hadronic weak decays of $D$
and $B$~mesons, copiously produced in $e^+e^-$~collisions above the
respective open-flavor thresholds,
allow the study of hadron resonances with well-defined quark flavors.

\subparagraph{Exotic meson production in glue-rich environments?}
Of particular interest in the search for exotic mesons are so-called
``glue-rich'' environments. These are experimental or physical situations in
which gluons are expected to play a dominant, active role in the dynamics,
rather than just (naively) binding quarks together. In most ordinary
hadrons, gluons act as the force mediators that hold quark–antiquark
pairs or three-quark systems together. However, in glue-rich
environments, the gluonic fields themselves are assumed to be highly excited or
abundant, increasing the likelihood of forming states where gluons
contribute directly to the particle’s valence structure and overall
spin-parity quantum numbers.

Historically, production of glueballs has mainly been predicted for glue-rich environments.
The most promising examples are proton-antiproton annihilation,
central $pp$ or $\pi p$ collisions through
double-Pomeron\,\footnote{In high-energy scattering theory, the Pomeron is associated
  with the exchange of vacuum quantum numbers (no electric charge, no
  color, no flavor, etc.), responsible for soft, diffractive processes.}
exchange or radiative decays of quarkonia (vector states), where the
spin of the state is transferred to the emitted photon,
leaving two gluons to form bound states. The
most studied example is the radiative decay $J/\psi\to \gamma G$.
\begin{enumerate}
  \item In $p\bar{p}$~annihilations, glueballs may be formed when
    quark-antiquark pairs annihilate into gluons. Though not very
    likely, this may proceed via formation (as opposed to production) without a recoil particle; in
    this case, exotic quantum numbers are forbidden, and the properties of
the glueball candidate can be determined from the initial state. Instead, production of a heavier resonance recoiling against another
meson is normally expected. However, experiments at the Low-Energy
Antiproton Ring (LEAR) at CERN
demonstrated that the
usually observed final states consisting of light $u$ and $d$ quarks can be just as effectively produced by
quark rearrangements, i.e., without glueballs in the intermediate
state. And no glueball has been unequivocally confirmed in $p\bar{p}$~annihilation.
\item The study of radiative decays of quarkonia is considered most suggestive in the glueball search. Most of
the information in this field has centered on $J/\psi$ decays; after photon emission, the $c\bar{c}$~annihilation can
go through C-even gg states and hence, may have a strong coupling to the low-lying glueballs. Study
of $J/\psi$~decays facilitates the search because the open-charm
$D\bar{D}$~threshold is above the $J/\psi$ mass of 3097~MeV
and the OZI rule\,\footnote{The OZI rule is based on an empirical
  observation that processes are suppressed if they require
  disconnected quark-line diagrams.} suppresses decays of the
$c\bar{c}$~system into light quarks. The study of radiative
$J/\psi$~decays was a major motivation for the CLEO-c experiment at
Cornell University, USA, and the BES experiment in Beijing, China. 
\item In contrast to formation experiments in $e^+e^-$ and $p\bar{p}$~annihilation,
the total energy in production experiments is shared among the recoiling particle(s) and the multi-meson
final state. The creation of new particles is limited by the center-of-mass energy of the reaction. The
mass and quantum numbers of the final state cannot be determined from the initial state, and thus,
many resonant waves with different angular momenta can contribute to the reaction; resonances are
not simply observed as structures in the cross section.

In central production, it was suggested that glueballs would be produced copiously in the process
\begin{equation}
    {\rm hadron}~{\rm beam}~+~p\to {\rm hadron}_f~X~p_s\,,
\end{equation}
where the final-state hadrons carry large fractions of the initial-state hadron momenta and are scattered
diffractively into the forward direction. To fulfill this requirement in a (proton) fixed-target experiment,
a slow proton and a fast hadron need to be observed in the final state. Mostly proton beams were
used for these kinds of experiments, but some also involved pion or
even Kaon beams.
\end{enumerate}  

\subsection{(Light)-meson systems}
According to the 2025 edition of the Review of Particle Physics,
the Meson Summary Table contains a total of 86 unflavored light mesons
$(C=S=B=0)$ and 28~strange mesons $(S=\pm1,C=B=0)$. Among these,
53~unflavored and 20 strange mesons are classified by the Particle
Data Group as established states. In addition, the RPP lists 30~charmonium-like $(c\bar{c})$ and 20 bottomonium-like
$(b\bar{b})$ resonances, together with 15~open-charm $(C=\pm 1)$ and
10~open-bottom $(B=\pm 1)$~mesons. Of these, 22~charmonium-like and
18~bottomonium-like resonances, as well as 9~open-flavor charmed and
8~open-flavor bottom states, are considered established by the
PDG. Furthermore, 11~charmed–strange $(C=\pm 1, S= \pm 1)$ and
7~bottom–strange $(B=\pm1,S=\pm 1)$ resonances are listed, of which
8~and~4, respectively, are regarded as established. More recently, two
bottom–charm mesons with $B=C=\pm 1$, namely the $B_c^+$ and $B_c(2S)^\pm$,
have been added to the spectrum. The observed conventional
heavy-flavor mesons with established quantum numbers are shown in
Table~\ref{chap2:tab1}. In parallel, a substantial number of
candidate tetraquark states has been reported in recent years.

In the following sections, a brief overview of the presently known
mesons and meson-like resonances with orbital angular momentum $L\leq 2$ is presented.
Beyond the $L=2$~sector, the spectroscopy of mesons becomes
considerably less established, with the available experimental
information being concentrated mainly on the states with the highest
total angular momentum~$J$. In the $L=3$~sector, candidates for the
$J^{PC} = 4^{++}$~nonet have been identified. In addition, the
presence of $2^{++}$, $3^{++}$, $3^{+-}$~nonets is expected.
For $L=4$~mesons, the highest-spin configuration corresponds to
$J^{PC} =5^{--}$, and several candidates with these quantum numbers
are known. One also expects additional $3^{--}$, $4^{--}$, and
$4^{-+}$~nonets, for which only a limited number of observations
have been announced. A detailed discussion of these
states with $L>2$ lies beyond the scope of this chapter of the
Encyclopedia of Nuclear Physics, and the reader is
referred to the literature for further
information~[\cite{ParticleDataGroup:2024cfk}].

\subsubsection{Pseudoscalar mesons}
The pseudoscalar mesons are spin singlets ($S=0$) with no internal orbital
angular momentum. Therefore, the total spin is $J=0$ and their parity is negative due to
opposite parities of quark and antiquark. The ground-state nonet of
light pseudoscalar mesons contains the $\pi, \eta,\eta^{\,\prime}$, and $K$~mesons (see
Table~\ref{chap2:tab1}) and is the best studied of the meson
multiplets. The properties of the ground-state pseudoscalar mesons have been
measured with very high accuracy in many decades of experiments. The
pion was discovered in 1947 using photographic emulsions exposed to
cosmic rays and confirmed the existence of the particle predicted by
Hideki Yukawa in 1935 to explain the nuclear force. The charged and
neutral pions are among the particles in the Standard Model that are
best understood. Likewise, the $\eta$ and $\eta^{\,\prime}$ isoscalar
partners of the pion have been extensively studied in a large variety of
decay channels, which include rare and forbidden processes. These are
sensitive to symmetry breaking effects and higher-order contributions in effective field theories. 
In contrast, excited light-pseudoscalar mesons are known less well. The PDG identifies
the $\pi(1300)$, $\eta(1295)$, $\eta(1475)$, and $K(1460)$ as members
of the nonet of radially excited pseudoscalar
mesons~[\cite{ParticleDataGroup:2024cfk}]. It is worth noting that the PDG
considers three excited $\eta$~mesons as being established, $\eta(1295)$,
$\eta(1405)$, and $\eta(1470)$, but also notes that the latter two mesons may be
manifestations of a single state. Some doubt has also been raised
about the existence of the $\eta(1295)$ that has always been observed
in conjunction with the $f_1(1285)$. Some feedthrough from this
$1^{++}$~state could possibly explain the isospin mixing. Other assignments of observed
states to multiplets are challenging. In the 1700--1800~MeV mass
range, the pseudoscalar glueball and an additional nonet of $J^{PC} = 0^{-+}$
hybrid mesons are also expected.

Heavy pseudoscalar mesons are the simplest bound states containing a
heavy quark, and they serve as a laboratory for studying how the
strong and weak forces interplay in real particles. Following the
naming scheme for light mesons, the isoscalar states with hidden charm
or bottom are called $\eta_Q$, where the index denotes the heavy quark
($Q = c,b$). These are roughly $c\bar{c}$ or $b\bar{b}$ states mixed
with light flavors.
The charmed triplet contains the $D^0$, $D^+$,
and $D^+_s$~mesons, whereas the charmed anti-triplet contains the
$\bar{D}^0$, $D^-$, and $D^-_s$~mesons, see
Fig.~\ref{chap2:fig1} (right side). Here, the index~$s$
denotes the strange quark. For the bottom multiplets, the {\it D} is replaced with a {\it B} in
the meson name. 

Despite the overall maturity, many questions still need to be addressed. Rare
decays, including those that violate conserved symmetries or are
highly suppressed in the SM, are being pursued with
increasing sensitivity, for example by the GlueX Collaboration at
Jefferson Lab. These measurements are motivated both by the
desire to test QCD in the non-perturbative regime and to search for
signs of new physics. Additionally, while the ground states are well
understood, the situation for excited pseudoscalar mesons is less
clear, with ongoing efforts to refine their identification and
properties. Particularly in the heavy-flavor sector, the PDG considers 
only $\eta_c(2S)$ a well-established state.

\subsubsection{Vector mesons}
The vector mesons are spin triplet ($S=1$) mesons with no internal orbital
angular momentum. Similar to the pseudoscalar mesons, their parity is negative due to
opposite parities of quark and antiquark, resulting in the overall
quantum numbers $J^{PC} = 1^{--}$. The nonet of light ground-state vector
mesons contains the $\rho, \omega,\phi$, and $K^\ast$~mesons (see
Table~\ref{chap2:tab1}). The dominant decay modes
of the vector mesons are through the strong interaction to
two or three pseudoscalar mesons, and the states are nearly
ideally mixed with the $\omega$ almost all $u\bar{u}$ and $d\bar{d}$,
whereas the $\phi$~meson is almost purely $s\bar{s}$. The
$\phi$~meson is particularly interesting due to its relatively narrow
width. Precision measurements of its decays into Kaon pairs and
radiative final states have been used to study SU(3)$_f$~symmetry
breaking and electromagnetic structure. The PDG identifies
the $\rho(1450)$, $\omega(1420)$, $\phi(1680)$, and $K^\ast(1410)$
mesons as members
of the nonet of radially excited vector mesons. Other excited states
are known, but may mix with a predicted nonet of $1^{-}$ hybrid
mesons. Disentangling these states requires high-statistics data and
careful modeling of interference effects in sophisticated amplitude
analyses.
Established orbital $D$-wave excitations ($L=2$) include $\rho(1700)$,
$\omega(1650)$, $\phi(2170)$, and
$K^\ast(1680)$~[\cite{ParticleDataGroup:2024cfk}]. However, the PDG
notes that the $\phi(2170)$~meson has also been proposed as a tetraquark
state or hybrid meson. Moreover, the physical state, $K^\ast(1680)$, is a mixture of
$1\,^3D_1$ and $2\,^3S_1$.  

The naming scheme for heavy vector mesons follows the nomenclature for
the heavy pseudoscalars in that the open-flavor mesons are assigned an
additional $^\ast$. The hidden-flavor states are called $J/\psi\,(1S)$,
$\psi(2S)$, $\psi(3S)$ and $\Upsilon\,(1S)$, $\Upsilon(2S)$, $\Upsilon(3S)$~states, etc. The charmed vector mesons are
certainly special in particle physics because the discovery of the $J/\psi$~resonance in
1974 at SLAC and BNL provided the first clear evidence for the
existence of a fourth quark flavor and, as such, key predictions of the
Standard Model. For this reason, and since it paired nicely with
existing names like ``strange,'' the new quark flavor was called {\it
  charm.} Heavy vector mesons can be fairly easily produced in
$e^+e^-$~collisions at center-of-mass energies above 3~GeV via the
production of a virtual photon, e.g., in the reaction $e^+
e^-\to\gamma^\ast\to c\bar{c}$. The large data sets of charmed vector
mesons are used to study light flavors and to search for exotic mesons
including glueballs. Several excited heavy-vector mesons are
established, and the open-charm states include $D^\ast(2007)^0$,
$D^\ast(2010)^\pm$, and $D_s^{\ast\pm}$. The radial excitations of the $J/\psi$ are
$\psi(2S)$, $\psi(4040)$, $\psi(4415)$, and $\Upsilon$~states are also
considered established by the
PDG up to $\Upsilon(4S)$. Two additional $1^{--}$~states are established,
$\Upsilon(10860)$ and $\Upsilon(11020)$. Orbital D-wave excitations
(L= 2) are known and established in the charm sector and include
$\psi(3770)$ and $\psi(4160)$, as well as
$D^\ast_{s1}(2700)^\pm$. However, the latter state is a mixture of  $1\,^3D_1$ and $2\,^3S_1$

\subsubsection{Pseudovector and axial vector mesons}
The pseudovector mesons are spin-singlet ($S=0$)~states with internal orbital
angular momentum ($P$-wave states). For this reason, their parity is
positive, $P = -(-1)^1 = +1$, and their $C$ parity negative, $C =
(-1)^1 = -1$, resulting in the overall
quantum numbers $J^{PC} = 1^{+-}$. The ground-state nonet of light pseudovector
mesons contains the $b_1(1235), h_1(1170), h_1(1415)$, and $K_{1B}$~mesons (see
Table~\ref{chap2:tab1}) with isospin $I=1, 0, 0, 1/2$, respectively. On the
other hand, the axial vector mesons are spin-triplet ($S=1$)~states, again with internal orbital
angular momentum ($P$-wave states). Thus, their parity is
positive, $P = -(-1)^1 = +1$, and their $C$-parity is also positive, $C =
(-1)^2 = +1$, resulting in the overall
quantum numbers $J^{PC} = 1^{++}$. The ground-state nonet of light
axial vector mesons contains the $a_1(1260), f_1(1285), f_1(1420)$, and $K_{1A}$~mesons (see
Table~\ref{chap2:tab1}). Note that both pseudovector and axial vector
states with open strangeness have $J^P = 1^+$ since the $C$-parity is not defined. For this reason,
the two states can mix and the two physical states, $K_1(1270)$ and
$K_1(1400)$, are mixtures of the SU(3) states, $K_{1A}$ and $K_{1B}$,
with a mixing angle of $\theta_{K_1} = -(33.6\pm
4.3)^\circ$~[\cite{ParticleDataGroup:2024cfk}]. Beyond the ground states, little is known
about excited states with only a few candidate states. The PDG does
not consider any excited pseudovector state
as being established and only one excited axial vector state, $a_1(1640)$, as
the radial excitation of the $a_1(1260)$.  

Heavy hidden-flavor pseudovector mesons are called $h_c(1P)$ and
$h_b(1P)$, and heavy hidden-flavor axial vector mesons are called $\chi_{c1}(1P)$ and $\chi_{b1}(1P)$ for
charmed and bottom mesons, respectively.
The nomenclature for
open-flavor pseudovector and axial vector mesons is the same. The
$I=0$ states containing a strange quark are assigned an index ``s1,'' and likewise, the
$I=1/2$~states are assigned an index ``1''. The pseudovector triplets and
anti-triplets contain $D_1(2420)$ and $D_{s1}(2536)^\pm$ in the
charmed sector, and $B_1(5721)$ and $B_{s1}(5830)^0$ in the bottom
sector. The axial vector triplets and
anti-triplets contain $D_1(2430)$ and $D_{s1}(2460)^\pm$ in the
charmed sector; no candidate states exist in the bottom sector. There
are also no established excited states.

\subsubsection{Scalar mesons}
Light scalar mesons deserve a special place
in the discussion of light mesons for many reasons. 
They are important for understanding chiral symmetry breaking in QCD, and the famous
$\sigma$~meson, now established as $f_0(500)$, is often interpreted as
a key player in the chiral symmetry breaking mechanism. 
A full discussion goes well
beyond the scope of this chapter of the Encyclopedia of Nuclear Physics, and thus, we refer to the
literature for a deeper discussion. Scalar mesons do not fit neatly into the
traditional quark model classification. Many known scalar mesons have
fairly low masses and are unusually broad, and therefore, many observed
states are good candidates for unconventional mesons. Predictions for light
glueballs, tetraquarks, and meson-meson molecules fall into the same mass
range: scalar mesons are prime candidates for glueball mixing. 

Conventional scalar mesons are considered spin triplet ($S=1$)~states with internal orbital
angular momentum ($P$-wave states). The total spin is $J=0$, their
parity is positive, $P = -(-1)^1 = +1$, and their $C$~parity is also positive, $C =
(-1)^2 = +1$, resulting in the overall quantum numbers $J^{PC} = 0^{++}$.
Considered as being established by the PDG, the conventional ground-state nonet of light scalar
mesons contains the $a_0(1450)$ and $K_0^\ast(1430)$~mesons (see
Table~\ref{chap2:tab1}), but three $f_0$~states serve as candidates
for the two isoscalar states of the nonet: $f_0(1370)$, $f_0(1500)$,
and $f_0(1710)$. A second group of lighter scalar mesons is
frequently considered to form a nonet:
the $a_0(980), f_0(980), f_0(500)$ (also known as $\sigma$), and
$K_0^\ast(700)$ (also known as $\kappa$). The existence of the latter
two states has been very controversial since they are very broad and
challenging to extract from data. As of the 2018 edition of the RPP,
these two states both appear in the Meson Summary Table and are considered
as being established by the PDG. The interpretation of this nonet
remains controversial, but many phenomenological approaches tend to converge 
on a non-ordinary quark-antiquark nature, with a likely predominant
meson-meson type component (meson-meson molecule, meson cloud, etc.). But mixing
with other states is also likely. We refer to the mini-review ``Note on Light
Scalars'' in the RPP~[\cite{ParticleDataGroup:2024cfk}] for more details
and a long list of references on the topic.

In the heavy-flavor sector, only the charmed ground-state scalar meson
{\bf 16}-plet is
considered as being established by the Particle Data Group. In
addition to the light-flavor nonet, the members are $\chi_{c0}(1P)$
(and $\chi_{b0}(1P)$ in the bottom sector),
$D_0^\ast(2300)$, and $D_{s0}^\ast(2317)^\pm$. The latter state has
also been suggested as a tetraquark or $D^{(\ast)}K$~molecule.

\subsubsection{Tensor mesons and pseudotensor mesons}
The tensor mesons are spin-triplet ($S=1$)~states with internal orbital
angular momentum ($P$-wave states). For this reason, their parity is
positive, $P = -(-1)^1 = +1$, and their $C$ parity is also positive, $C =
(-1)^1 = +1$, resulting in the overall
quantum numbers $J^{PC} = 2^{++}$. The ground-state nonet of light tensor
mesons contains the $a_2(1320), f_2(1270), f_2^{\,\prime}(1525)$, and $K_2^\ast(1430)$~mesons (see
Table~\ref{chap2:tab1}). This nonet is almost ideally mixed. Also similar to the
situation of the vector mesons, an additional $L,S$~combination can
couple to $J^{PC} = 2^{++}$, i.e., $L=3$ and $S=1$, rendering the
assignment of physical states to nonets more difficult. The lightest
tensor glueball is expected to fall into the mass range of radial
excitations. The PDG identifies the $a_2(1700), f_2(1640), f_2(1950)$,
and $K_2^\ast(1980)$ as members of the nonet of radially excited
tensor mesons, but the $f_2(1950)$ has also been proposed as the
ground-state tensor glueball. The situation for tensor mesons is confusing. No
additional $I=1$ and $I=1/2$~states are known, but a zoo of
$f_2$~states has been reported. Likely, several states are the same
or artifacts, states with $L>1$, or glueball candidates. 

The pseudotensor mesons are spin-singlet ($S=0$)~states, with internal orbital
angular momentum ($D$-wave states). Thus, their parity is
negative, $P = -(-1)^2 = -1$, and their $C$-parity is positive, $C =
(-1)^2 = +1$, resulting in the overall
quantum numbers $J^{PC} = 2^{-+}$. The ground-state nonet of light
pseudotensor mesons contains the $\pi_2(1670), \eta_2(1870), \eta_2(1645)$, and $K_2(1770)$~mesons (see
Table~\ref{chap2:tab1}). Very little is known about excited states. Radial
excitations and a nonet of hybrid mesons are expected. The PDG
considers only one additional state established, $\pi_2(1880)$,
although some other candidates have been observed.

In the heavy-flavor sector, the naming scheme for tensor mesons is
similar to the naming scheme for scalar and axial vector mesons in
that the index contains the total $J$, see
Table~\ref{chap2:tab1}. The singlet states are $\chi_{c2}(1P)$ and
$\chi_{b2}(1P)$, whereas the triplet and anti-triplet states are
$D_2^\ast(2460)$ and $D^\ast_{s2}(2573)^\pm$, as well as
$B_2^\ast(5747)$ and $B^\ast_{s2}(5840)^0$. No heavy-flavor
pseudotensor mesons are known.

\subsection{Exotic Mesons}
The search for mesons with exotic quantum numbers is one of the most
direct and conceptually important tests of QCD. At stake is a
deceptively simple question: Are quarks the only building blocks of
hadrons, or do gluons also participate as active constituents in bound states?
Many of the exotic-meson configurations have the same quantum numbers as
ordinary mesons and can mix with them, rendering their identification
very challenging. However, some exotic mesons have quantum numbers that cannot be
produced by a simple quark–antiquark system, making their observation clear
signatures of a non-standard structure. Over the past two decades,
experiments have reported many candidates for exotic mesons, and the
most prominent examples are
the so-called X, Y, and Z states in the heavy-flavor sector.
These discoveries have opened a new window into how quarks and gluons
combine, revealing that the spectrum of hadronic matter is richer and more intricate than once thought.
In a brief summary, recent experiments in the light- and heavy-flavor
sectors have provided compelling experimental evidence for the
existence of exotic mesons.
\begin{BoxTypeA}[chap2:box1]{Overall, exotic mesons can be grouped into three broad categories:}
\begin{enumerate}
  \item Spin-exotic states, which have $J^{PC}$~quantum numbers that
    are not allowed or possible for ordinary mesons, the quantum
    numbers of which are based on the simple
    $q\bar{q}$~structure. These ``exotic'' quantum numbers are
    $0^{--}, 0^{+-}, 1^{-+}, 2^{+-}$, etc. Hybrid mesons, and oddballs (glueballs with exotic quantum
    numbers), or other non-standard configurations fall into this
    category. Note that the lightest glueball with exotic quantum
    numbers is predicted in the mass range 3--4~GeV, far beyond all known light-flavor
    mesons.
  \item Flavor-exotic states, which have flavor quantum numbers, such
    as isospin or strangeness, that are not possible for ordinary
    $q\bar{q}$~states.
  \item Crypto-exotic states, which have quantum numbers of
    conventional $q\bar{q}$~state and therefore, they can mix with these
    ordinary states. A very prominent example is the decades-old search for
    the scalar glueball. But also hybrid mesons,
    ($q\bar{q}q\bar{q}$) tetraquarks, and $(q\bar{q})(q\bar{q})$
  molecular states can
    have non-exotic quantum numbers.
  \end{enumerate}
\end{BoxTypeA}
Crypto-exotic states are very
    challenging to identify. Analyses require high-statistics data
    samples, and since states typically do not show up as clear structures 
in mass distributions or cross sections, 
    sophisticated amplitude analysis tools are needed. States can
    then be identified as supernumerary states in the regular
    multiplet picture. For this reason,
 the identification of exotic quantum numbers
 that are not possible for ordinary mesons is considered the most
 straightforward approach to unambiguously establish an exotic
 state. Roughly, the seach for exotic spin-parity $J^{PC}$ quantum
 numbers is performed in the light-flavor sector, whereas flavor
 exotic states with a minimum quark content of four quarks are
 reported from the heavy-flavor sector.

The search for glueballs falls into the realm of crypto-exotic states
since the lightest glueballs with exotic quantum numbers are predicted
at very large masses. The history of glueball searches is exciting and
unfortunately, the full story would go beyond the scope of this
chapter of the Encyclopedia of Nuclear Physics. We refer to the literature for all the glueball birth
announcements and obituaries. In a very brief summary, a prominent
glueball candidate
emerged in antiproton-nucleon experiments, $f_0(1500)$, leaving a
field of supernumerary $f_0$~states. In particular, the trio of
$f_0(1370), f_0(1500), f_0(1710)$ played an important role, and
attempts were made to interpret these states as the result of a mixing scheme
of the two expected isoscalar states with the scalar glueball. However, the
unusually broad $f_0(1370)$ decaying strongly into $\rho\rho$ could also
be interpreted as a molecular state. Additional scalar, isoscalar
states have since emerged in radiative $J/\psi$~decays, and the scalar
glueball is likely distributed over a larger mass range. Moreover,
very weak evidence for a tensor
glueball has been discussed, but similar to the scalar glueball, the
tensor glueball is likely also distributed over several tensor mesons.
The BES Collaboration has announced the observation of a pseudoscalar
glueball-like particle in 2024, but this state has yet to be confirmed
by another experiment~[\cite{BESIII:2023wfi}]. Identifying
a glueball remains a huge challenge.

\begin{figure}[t]
  \centering
  \begin{tabular}{ccc}
    \includegraphics[width=0.30\textwidth]{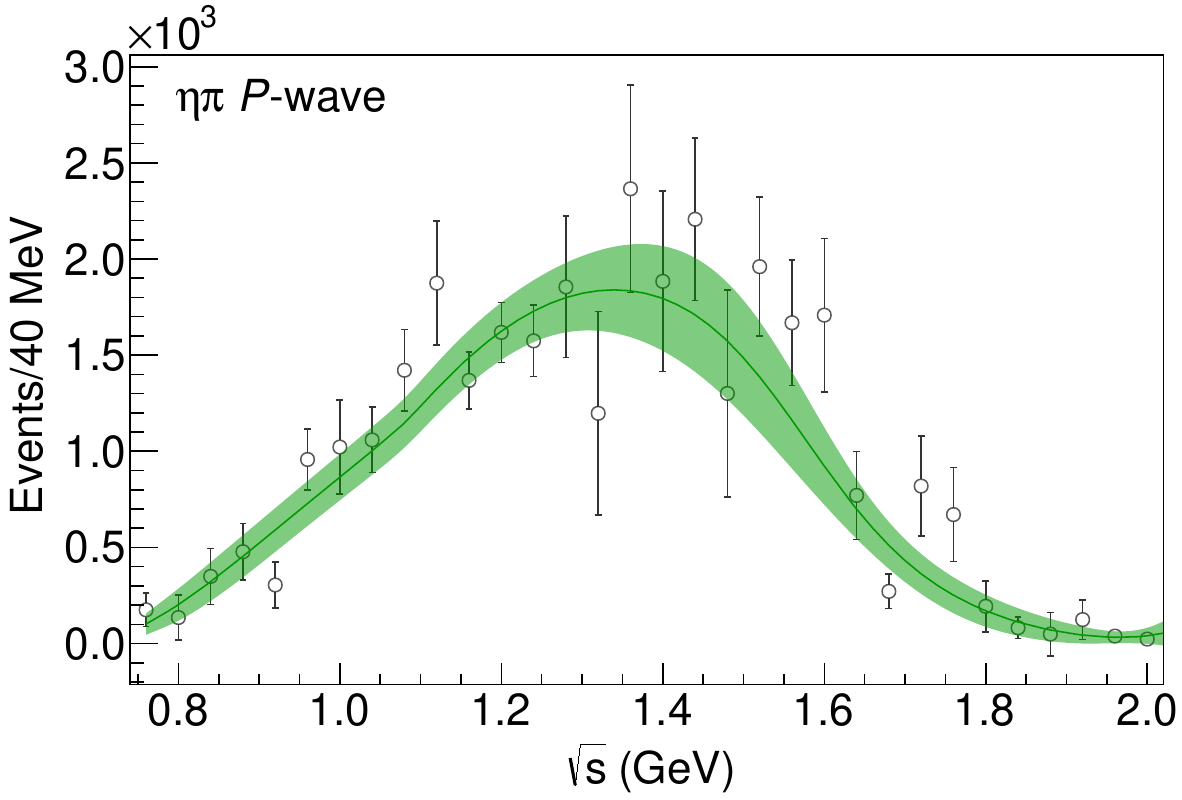} &
    \includegraphics[width=0.30\textwidth]{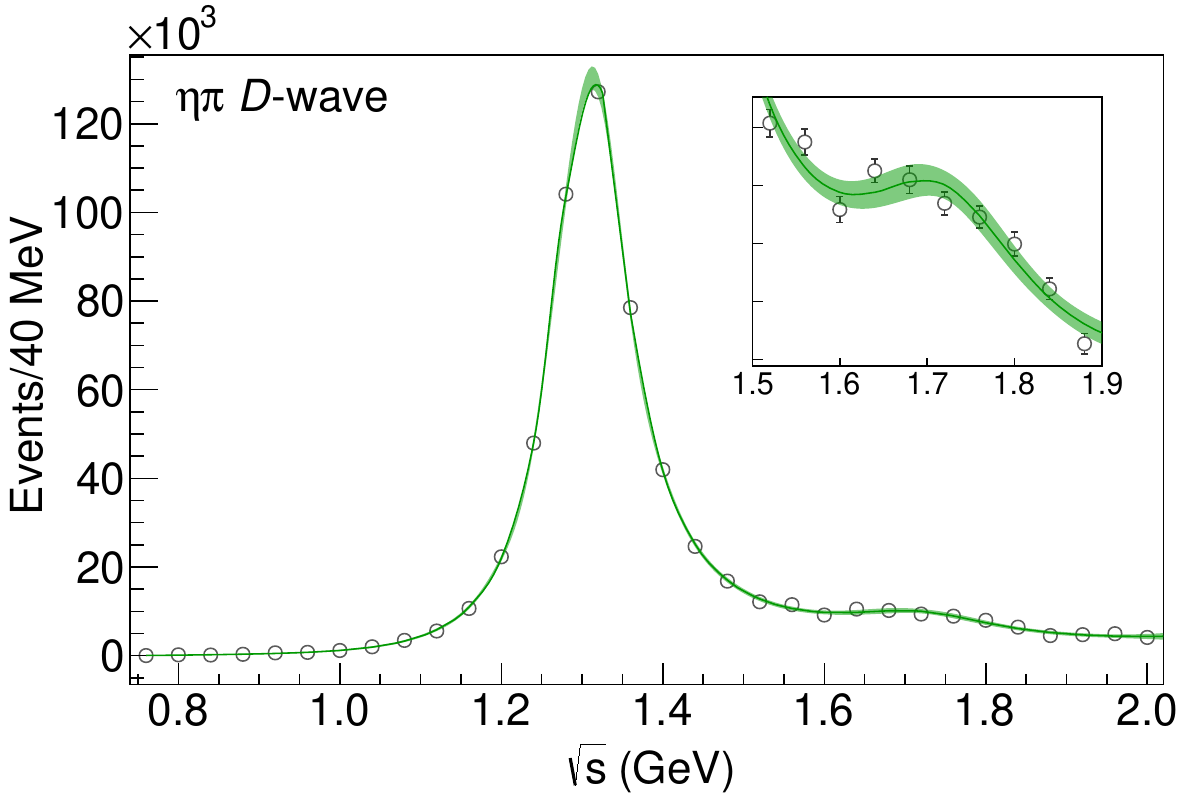} &
    \includegraphics[width=0.30\textwidth]{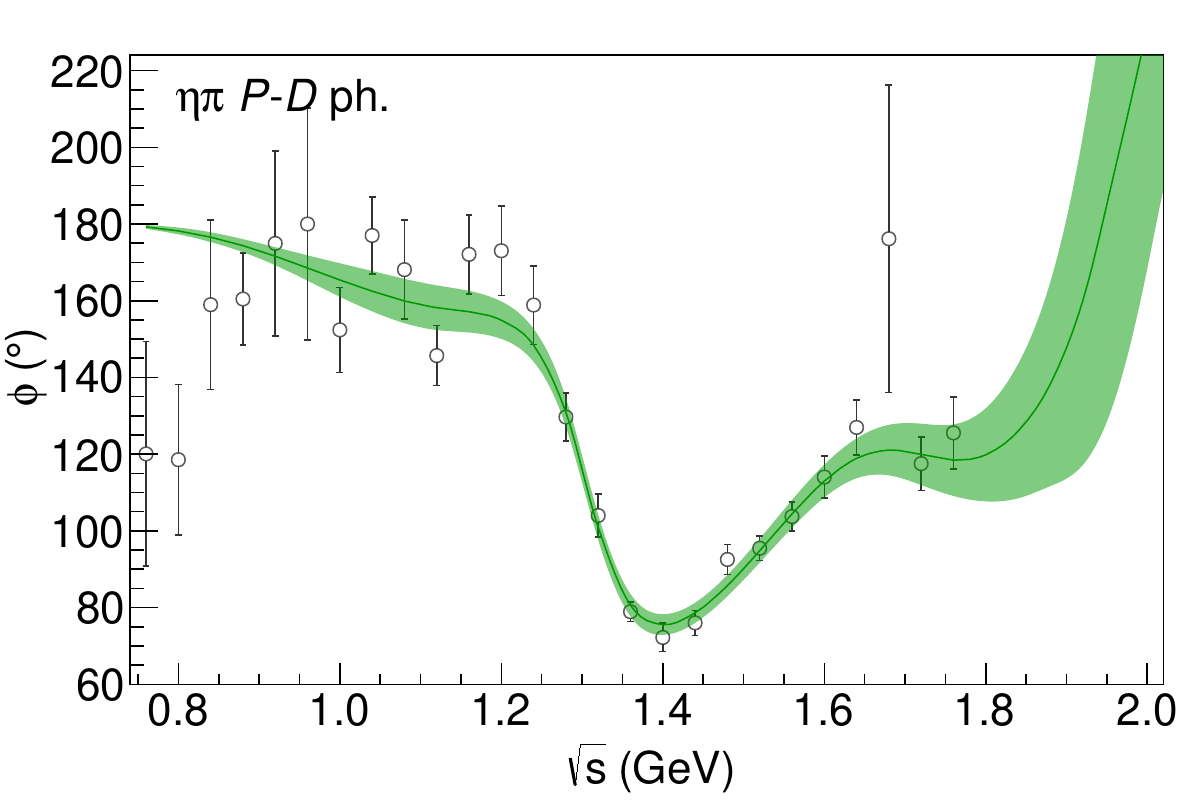}\\
    \includegraphics[width=0.30\textwidth]{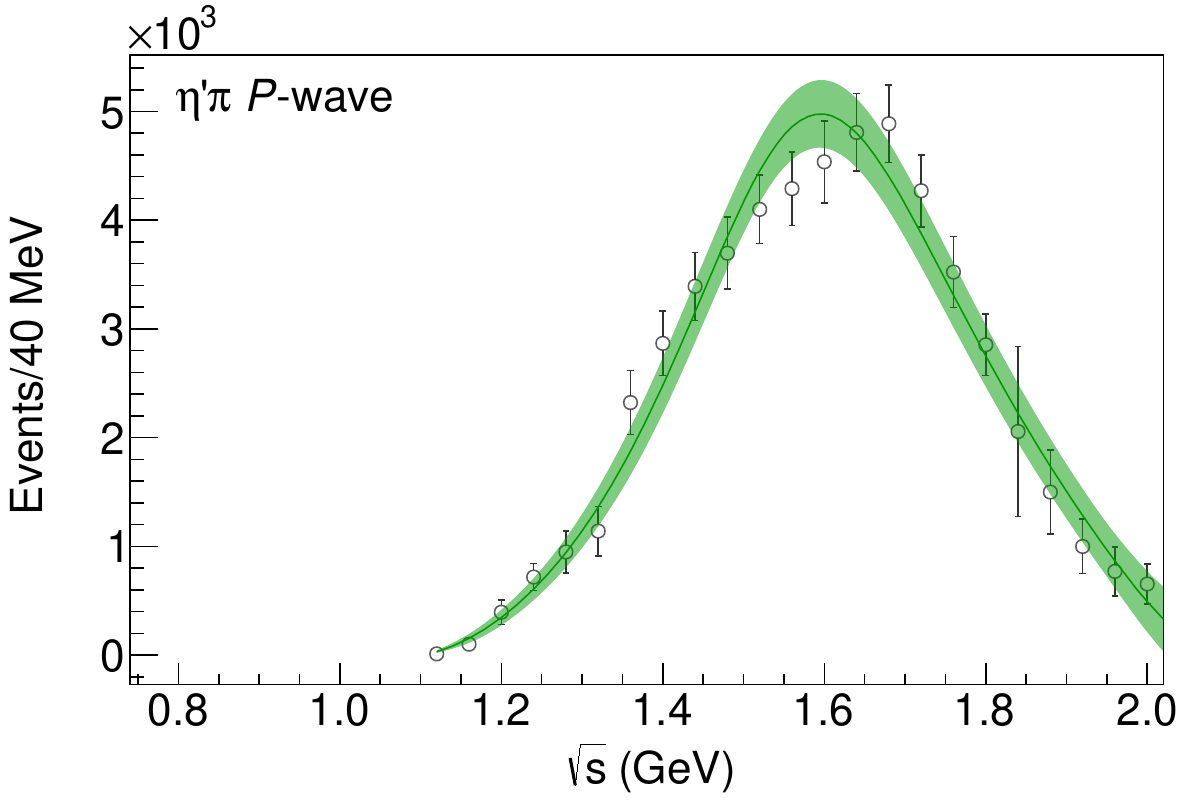} &
    \includegraphics[width=0.30\textwidth]{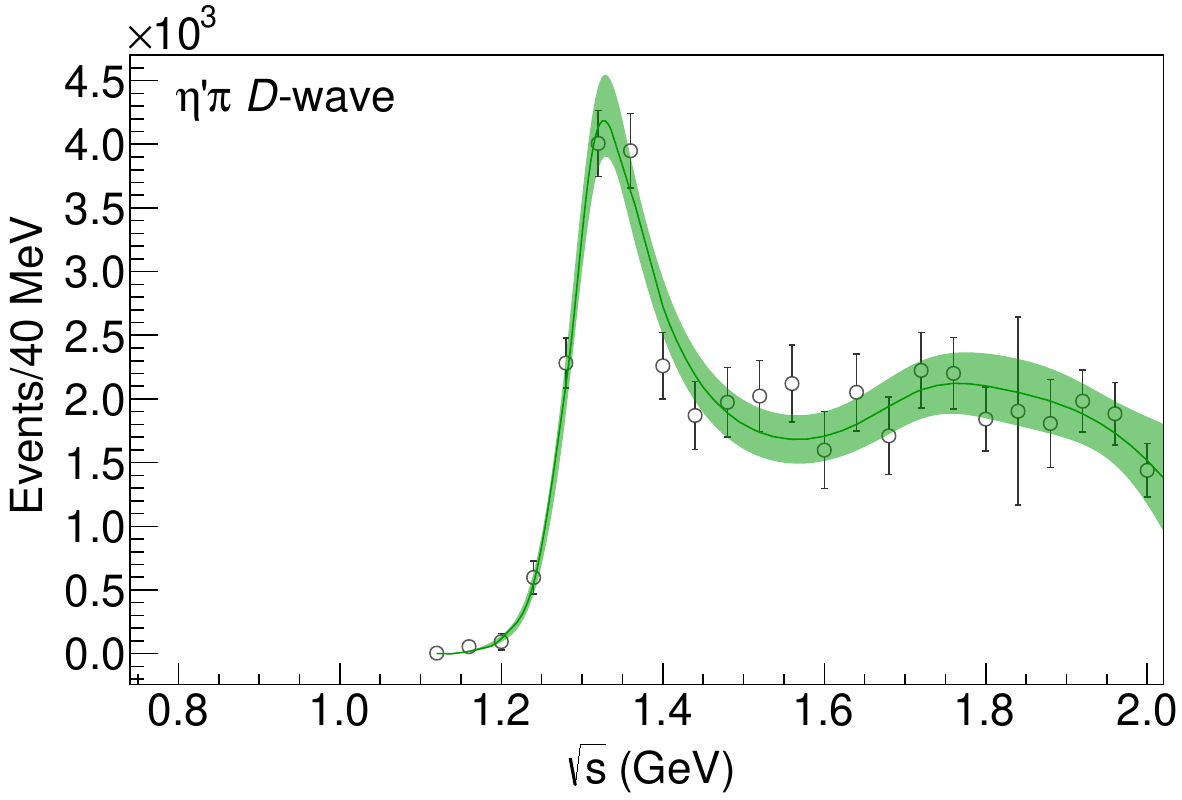} &
    \includegraphics[width=0.30\textwidth]{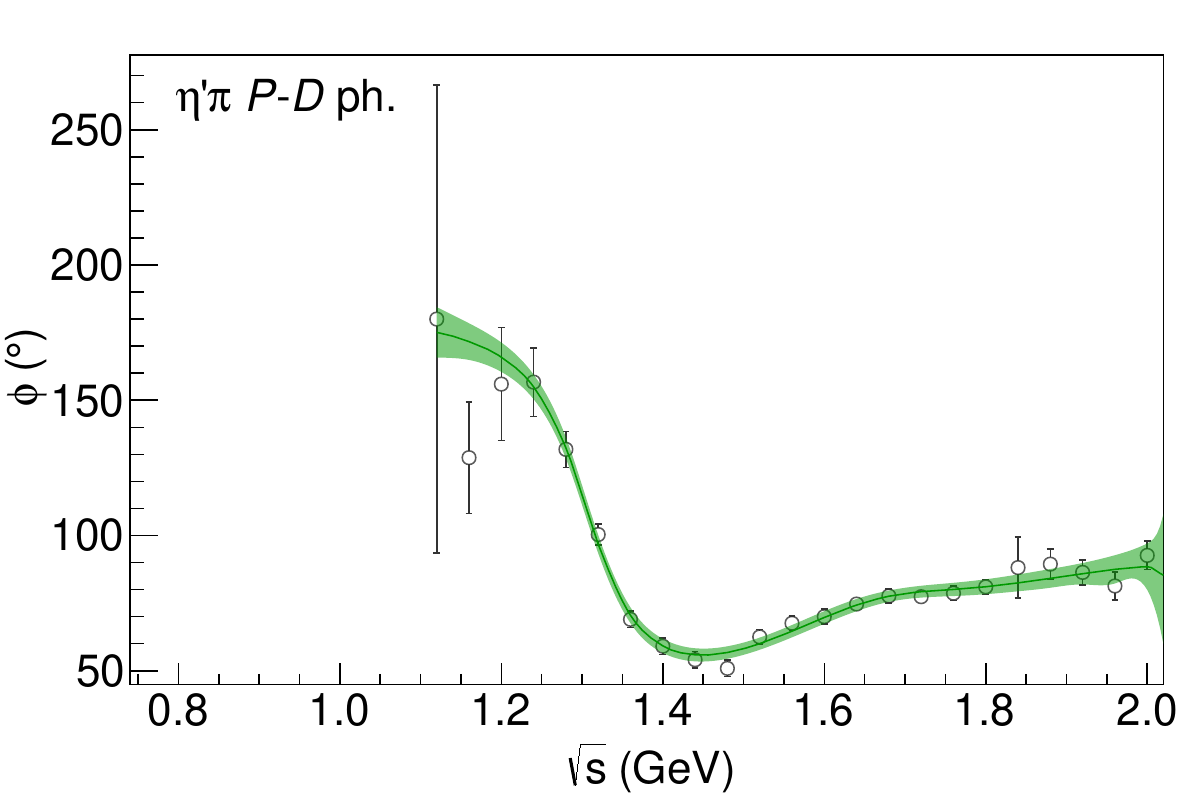}
  \end{tabular}
  \caption{Analysis by the JPAC Collaboration~[\cite{JPAC:2018zyd}]
    and fits to the $\eta\pi$ (top row) and $\eta^{\,\prime}\pi$
    (bottom row) data from COMPASS~[\cite{COMPASS:2014vkj}] that demonstrate that
    both exotic $\pi_1$~candidates can be described by a single
    resonance pole in the $\eta^{\,(\prime)}\pi$ $P$-waves. Shown are 
    the intensities of $P$- (left), $D$-wave (center), and their
    relative phases (right);
    the inset zooms into the region of the $a_2(1700)$. The
solid line and green band show the result of the fit and the
$2\sigma$~confidence level provided by the bootstrap analysis,
respectively. The best fit has $\chi^2/\text{dof} = 162/122=1.3$. The uncertainties shown are statistical
only. Reprinted figure with permission from~[\cite{JPAC:2018zyd}], Copyright (2019) by the American Physical Society.\label{chap2:fig2}}
\end{figure}

\subsubsection{Search for light spin-parity exotic mesons}
The experimental search for light exotic mesons has focused on the identification of
spin-exotic structures. Most models predict the lightest spin-exotic state
to be a hybrid meson with $J^{PC} = 1^{-+}$ quantum numbers in the
mass range 1.3--2.2~GeV. And the dominant decay modes for such a hybrid
meson are expected to be the $f_1(1285)\pi$ and
$b_1(1235)\pi$~channels. Other decay modes are more uncertain, and
model predictions diverge. The lattice spectrum published by the
Hadron Spectrum Collaboration~[\cite{Dudek:2011bn}] is qualitatively
similar to the one obtained from quark-model calculations.
However, the lattice calculation revealed an additional 
supermultiplet of extra states that lie about 1.3 GeV
above the lightest $J^{PC} = 1^{--}$~state and that have quantum
numbers of $0^{-+}, 1^{--}, 2^{-+}$, and $1^{-+}$, where the latter
multiplet is the famous nonet of spin-exotics.

Currently, the Particle Data Group
lists two spin-exotic light-meson states, $\pi_1(1400)$ and
$\pi_1(1600)$, based on a seemingly large body of evidence, including data from
pion diffraction, antiproton–nucleon annihilation, photoproduction,
and charmonium decays covering several decay
channels. First announced in 1988 by the GAMS Collaboration, more
recent high-precision data from the COMPASS Collaboration and more
advanced analysis techniques present a picture that is converging on a
single $\pi_1$ state. Figure~\ref{chap2:fig2} shows the results of a
coupled-channel analysis performed by the JPAC Collaboration~[\cite{JPAC:2018zyd}] using
$P$- and $D$-wave amplitudes for the $\eta\pi$ and
$\eta^{\,\prime}\pi$~channels measured by COMPASS~[\cite{COMPASS:2014vkj}]. Using only a single
resonance pole\,\footnote{A ``resonance pole'' refers to the location
  in the complex energy or momentum plane, where each pole typically
              corresponds to a physical feature such as a bound
              state, resonance, or virtual state.}
in the $\eta^{\,(\prime)}\pi$ $P$-waves, with pole parameters
consistent with those measured for the $\pi_1(1600)$, the JPAC model
was able to successfully describe both the $\eta\pi$ and
$\eta^{\,\prime}\pi$ partial-wave amplitudes. 

Despite various ambiguities, the $\pi_1(1600)$ has had a lasting impact on
the field. It has also motivated extensive theoretical work, from
phenomenological models of hybrid mesons to first-principles
calculations using lattice QCD. In this sense, its importance does not
depend solely on whether it is ultimately confirmed as a distinct
resonance. Rather, it has served as a focal point for efforts to
understand how QCD organizes the spectrum of hadrons beyond the
simplest quark model. The $\pi_1(1600)$ is emblematic of both the promise
and the difficulty of modern hadron spectroscopy. It embodies a
tantalizing possibility -- the direct observation of an exotic meson with
possibly gluonic degrees of freedom -- while simultaneously illustrating
the challenges of extracting clear physical conclusions from complex
strong-interaction data. Whether it is ultimately established as a
true hybrid meson or reinterpreted as a manifestation of more subtle
dynamics, its role in advancing our understanding of QCD is already
secure. In addition, it is important to search for the exotic
SU(3)$_f$~partner states of the $\pi_1$. Here, the 2022 announcement by the
BES\,III Collaboration of the possible observation of an
$\eta_1(1855)$~isospin-singlet $1^{-+}$~spin-exotic state in $J/\psi\to \gamma\eta\eta^{\,\prime}$
could be a major breakthrough~[\cite{BESIII:2022riz}]. But this state has yet to be confirmed
by another experiment or in a different channel. Last but not least, the search
for states with other spin-exotic $J^{PC}$~quantum numbers such
as $0^{+-}$ and $2^{+-}$ continues. These searches will also yield a
more complete picture of the spectrum of states with ordinary
quantum numbers, which not only helps identify supernumerary
states but is also an important input to theory in order
to improve our understanding of the non-perturbative regime
of QCD.

\begin{figure}[t]
  \centering
    \includegraphics[width=0.95\textwidth]{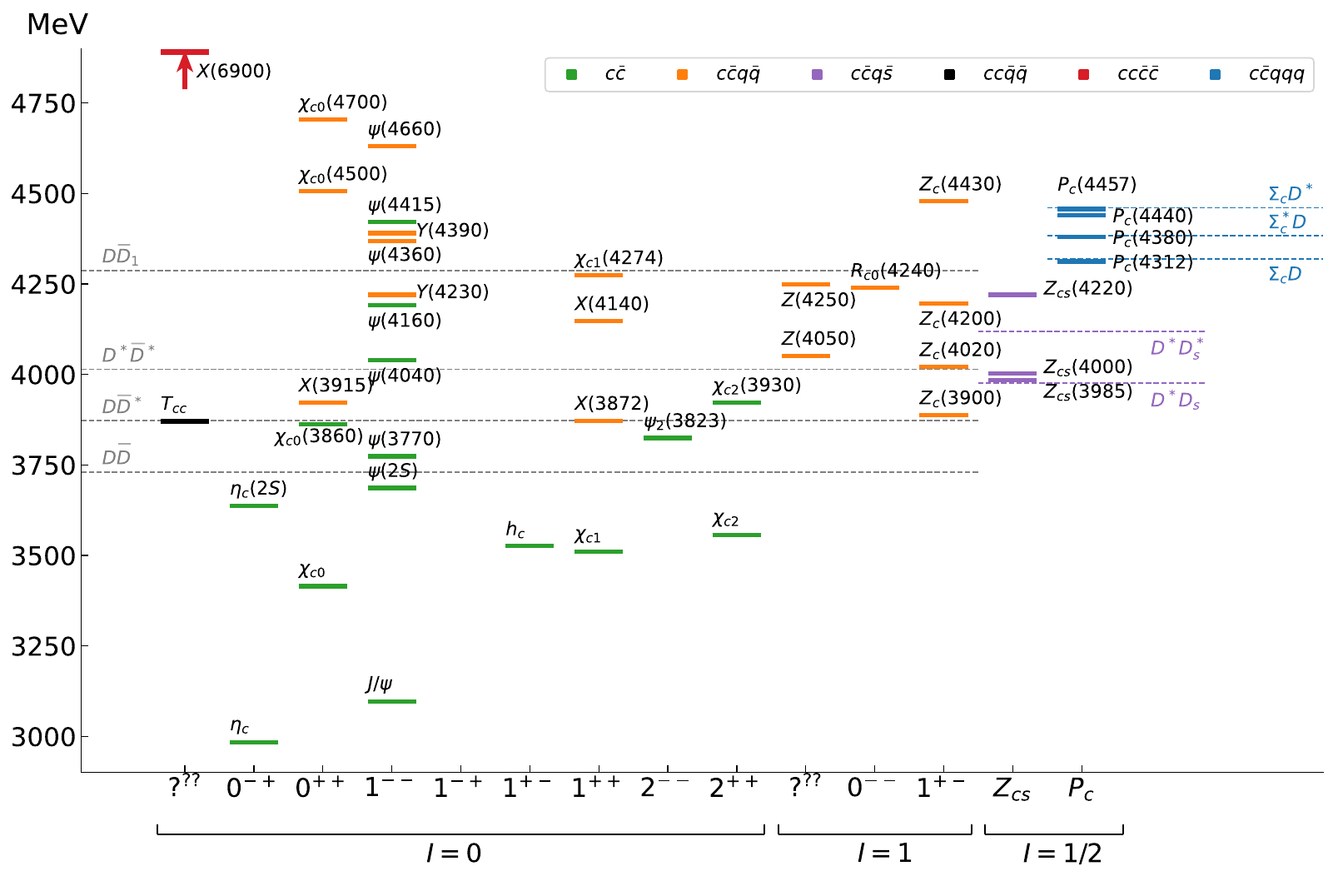}
  \caption{Status of charmonia, charmonium-like, and pentaquark states~[\cite{JPAC:2021rxu}]:
    States below the $D\bar{D}$~threshold as well as $\psi(3770)$,
    $\psi(4040)$, $\psi(4160)$, $\psi(4415)$,
    were discovered before 2003, whereas all other states were
    discovered after 2003. Reprinted from~[\cite{JPAC:2021rxu}], Copyright (2022), with permission from Elsevier.\label{chap2:fig3}}
\end{figure}

\subsubsection{Search for heavy flavor-exotic mesons}
In spectroscopy, mesons containing one or more heavy quarks are
located at the QCD sweet spot of the perturbative and non-perturbative
regime, where the strong interactions are both
rich and controllable. Heavy quarks provide a mass
scale that is bigger than $\Lambda_{QCD}$, and therefore, perturbation
theory is applicable and a valid tool. Moreover, heavy
quarkonia, $Q\bar{Q}^{\,\prime}$ with $Q,Q^{\,\prime} = c,b$, can be
treated non-relativistically since the quark velocities are small. These heavy states provide a relatively clean
laboratory for studying confinement and the quark–antiquark potential.

The search for exotic mesons has centered on light-flavor mesons for
many decades, but it is the heavy-flavor sector now where QCD exotics
have revealed themselves, and the existence of exotic mesons is
definitely established. From about 2003, new states above the
$D\bar{D}^{(\ast)}$ and $B\bar{B}^{(\ast)}$~thresholds were observed
and the discovery of the hidden-charm $X(3872)$~state by the Belle
Collaboration, as well as the open-charm $D_{s0}^\ast(2317)$~state by
the BaBar Collaboration, in particular, ushered in a new era in hadron
spectroscopy. These observations were followed by many other
structures, so-called $XYZ$~states. Figure~\ref{chap2:fig3} shows the
current spectrum of experimentally observed charmonia and charmonium-like
states.
Many of these new states have properties that do not
match those of conventional quarkonium states. For example, the mass
of the $X(3872)$ does not match expectations for predicted charmonium
levels, and the $D_{s0}^\ast(2317)$ is found as an extremely narrow
resonance below the $DK$~threshold. Instead, quark models predicted
the lowest scalar $c\bar{s}$~state ($J^P = 0^+$) to lie well above the
$DK$~threshold. The discussion turned quickly to the possibly exotic
structure of the growing number of $X$, $Y$, and $Z$~states.

While ``$X$~state''
was an early label for states whose nature was unclear, $Y$ and
$Z$~states have been more specifically defined. $Y$-states are usually
vector states with $J^{PC} = 1^{--}$ produced directly in $e^+e^-$
annihilation, and $Z$-states are ``charged'' hidden-charm or hidden-bottom
states, which must contain at least four quarks. Since their original
discovery, the properties of some of these states have become better
understood, and it has become possible to include some of them in the RPP
Listings as well as Summary Tables following standard naming
conventions. From the 2018 edition of the RPP, the above
$X(3872)$~state has now appeared as $\chi_{c1}(3872)$ since measurements
established its quantum numbers as $I^G\,(J^{PC}) = 0^+\,(1^{++})$. Note that the
name of a resonance is not related to the internal structure but
simply follows the convention to assign states to $J^{PC}$~multiplets. Electrically
charged charmonium-like $Z$~states with a four-quark content are now
called $T_{\rm \,quarks\,J}^{(\ast)}$~states, where the quark content
is explicitly listed as a subscript. Similarly to the naming scheme
for mesons, the spin $J$ is appended as a subscript after the quark
content. When the spin-parity is in the natural series
($J^P=0^+,1^-,2^+$, etc.), a superscript ``$\ast$'' is added.

In the 2025 edition of the RPP~[\cite{ParticleDataGroup:2024cfk}], a total of
19 $T$~states is listed in the Meson Summary Table under ``OTHER,'' and
six of these are considered as being established.
While it is still not possible to rule out firmly a conventional
nature for the majority of $X$, $Y$~states, the observation of
the $Z_c(4430)^+$~meson (now known as $T_{c\bar{c}1}(4430)^+$), an
electrically charged charmonium-like state, and of the
$T_{cc}(3875)^+$~state, a meson containing two charm quarks,
established the existence of QCD resonances of manifestly exotic nature.

\section{Spectroscopy of Baryon Systems: An Experimental Overview
  \label{BaryonSpectroscopy:sec3}}
Baryons play a central role in modern particle and nuclear physics,
not only because they constitute nearly all visible matter in the
universe, but also because they provide a unique laboratory for
exploring the dynamics of the strong interaction. Protons and neutrons
form the building blocks of atomic nuclei, while their excited states
and strange counterparts are believed to have played a crucial role in
the evolution of the early universe. Roughly $10~\mu$s after the Big Bang,
hadronic matter emerged from a quark-gluon plasma, and a realistic
description of this transition requires the inclusion of the full
spectrum of baryon resonances in hadron resonance gas models. Such
models are essential for understanding the freeze-out behavior
observed in hot-QCD calculations and for locating the critical point
associated with the transition between the quark-gluon plasma and
hadronic phases, expected near a temperature of 155 MeV. These
questions continue to motivate extensive experimental programs at
CERN, RHIC, and the future FAIR facility, where the phase diagram of
strongly interacting matter is investigated through heavy-ion
collisions over a wide range of energies.

At a more fundamental level, baryons provide the simplest systems in
which the {\it non-Abelian nature} of quantum chromodynamics becomes
manifest. In brief, a non-Abelian theory is characterized by symmetry
operations that do not commute. In QCD, this implies that the force
carriers of the strong interaction, the gluons, themselves carry color
charge and can therefore interact directly with one another. This
self-interacting structure gives rise to some of the most distinctive
features of QCD, including confinement, asymptotic freedom at high
energies, and the rich spectrum of hadron resonances. Furthermore, the proton, composed of three valence quarks,
directly reflects the existence of the three color degrees of freedom
underlying QCD. At the same time, the internal structure of baryons is
sufficiently rich to reveal aspects of strong-interaction dynamics
that remain hidden in mesonic systems. During the early development of
particle physics, mesons were not yet uniquely identified as
quark–antiquark states, and a variety of alternative interpretations
were explored, including elementary particles, composite hadronic
systems, and other symmetry-based constructions. However, the three-quark structure of baryons naturally led to
their successful organization into SU(3)-flavor multiplets,
culminating in the prediction and subsequent discovery of the
$\Omega^-$~baryon\,\footnote{The discovery of the $\Omega^-$~baryon in
1964 marked a major triumph for SU(3)$_f$~flavor symmetry. The
organization of hadrons into multiplets led to the prediction of a
missing baryon in the baryon decuplet with strangeness $S = -3$,
electric charge $-1$, and a mass near 1670~MeV.}. The study of baryons, therefore, provides not only
insight into the structure of matter itself but also a powerful
framework for understanding confinement, symmetry, and the emergent
phenomena of strong-coupling QCD.

\subsection{Baryon Classification\label{BaryonClassification:subsec1}}
Similar to mesons, baryons are classified mainly by their quark
content, spin, parity, and how they fit into spin-flavor symmetry
groups. Light-flavor, non-strange baryons are labeled $N$ and $\Delta$, whereas
light baryons that
contain at least one strange quark (strangeness $-1$) are called
hyperons and are labeled $\Lambda~(I=0)$, $\Sigma~(I=1)$ 
(with strangeness $S = -1$), $\Xi~(I=1/2)$ (with strangeness $S = -2$), and
$\Omega~(I=0)$ (with strangeness $S = -3$). Accordingly, antihyperons have
strangeness $+1$, $+2$, and $+3$. 
While for charged mesons the
positively and negatively charged states are typically antiparticles of each
other\,\footnote{Neutral mesons are more
subtle. While the $\pi^0$~meson is its own antiparticle, the
$K^0$ is distinct from the $\bar{K}^0$. Neutral Kaons are especially
important in particle physics because the $K^0$ and $\bar{K}^0$ can
transform into each other through weak interactions, leading to
phenomena such as Kaon mixing and CP violation.}, a common mistake is
to assume that this is also true for
baryons. For example, the $\Delta^+$ and $\Delta^-$~baryons are not
antiparticles of each other. Baryons consist of three valence quarks;
consequently, the corresponding antibaryons are obtained by replacing
each quark with its corresponding antiquark. Finally, for individual states within an
isospin multiplet, the electric charge is specified as a
superscript. In the case of isospin-zero $(I=0)$ states, such as the $\Lambda$ and $\Omega$~baryons,
the charge label is unnecessary because the states are uniquely
defined, although it may still be included. The mass of the baryon is
given in parentheses, and the spin-parity~$J^P$ is included as a
suffix. The positively charged $\Delta$~baryon may serve as an example:
$\Delta(1232)^+\,\frac{3}{2}^+$.

The naming scheme for light-flavor baryons
can be expanded to include heavy baryons, which are labeled with an
index. For example, the $\Lambda_c^+$ has isospin zero 
and quark content $|udc\rangle$, $\Xi_c^+$ has isospin $\frac{1}{2}$ and quark content $|usc\rangle$, and 
$\Xi_{cc}^{++}$ has isospin $\frac{1}{2}$ and quark content
$|ucc\rangle$, for instance. Historically, the term {\it hyperon}
refers specifically to baryons that contain at least one strange quark
but no charm or bottom quarks. Therefore, a baryon with quark content $|usc\rangle$ is
generally classified as a {\it charmed strange baryon} and not as a
traditional hyperon. Exotic baryons include pentaquarks with a $|qqq\bar{q}q\rangle$
quark content and are assigned the basename $P_{\rm
  quark~content}\,{\rm (mass)}^{\rm \,charge}$ followed by the spin-parity
$J^P$~quantum numbers.

\subsubsection{Light-flavor baryons\label{BaryonClassification:LightFlavors}}
Baryons are fermions, and therefore, they obey the Pauli exclusion
principle. Their total wave function
\begin{eqnarray}
|qqq\rangle_A = |{\rm color}\rangle_A\,\times\, |{\rm space,~spin,~flavor}\rangle_S
\end{eqnarray}
must be antisymmetric (denoted by the index~$A$) under the interchange
of any two equal-mass quarks. Note that 
all hadrons are color singlets, and for this reason, the color component of the wave function must be completely
antisymmetric. In the light-flavor baryon sector, SU(3)$_f$~symmetry
approximately holds for exchanges among the three quark
flavors $u$, $d$, and $s$, though it is broken by the higher mass
of the strange quark.
The flavor wave functions of these baryon states can then be
constructed as members of SU(3)$_f$~multiplets:
\begin{equation}
  {\bf 3}\,\otimes\, {\bf 3}\,\otimes\, {\bf 3}\,=\,{\bf 10}_S\,\oplus\, {\bf 8}_M\,\oplus\, {\bf 8}_M\,\oplus\,
  {\bf 1}_A\,,
\end{equation}
where the decuplet is fully symmetric under exchange of any two
quarks, the singlet is totally antisymmetric in flavor, and the octets
exhibit mixed symmetry, i.e., they are neither fully symmetric nor
antisymmetric.

The simple quark model based on three symmetric quark degrees of
freedom allows for two possible values of the intrinsic baryon spin,
either $S = \frac{1}{2}$ or $S = \frac{3}{2}$, where the spin wave
function exhibits a mixed symmetry or is totally symmetric,
respectively. Flavor and spin degrees of freedom can be combined into
an approximate spin–flavor SU(6) symmetry, whose multiplets are
\begin{eqnarray}
{\bf 6}\,\otimes\, {\bf 6}\,\otimes\, {\bf 6}\,=\,{\bf 56}_S\,\oplus\,
{\bf 70}_M\,\oplus\, {\bf 70}_M\,\oplus\, {\bf 20}_A\,.
\end{eqnarray}
These decompose into irreducible flavor-SU(3) multiplets
\begin{align}
{\bf 56}~=&~~ ^4{\bf 10}\,\oplus\, ^2{\bf 8}\\\label{Equation:70plet}
{\bf 70}~=&~~ ^2{\bf 10}\,\oplus\, ^4{\bf 8}\,\oplus\, ^2{\bf 8}\,\oplus\, ^2{\bf 1}\\\label{Equation:20plet}
{\bf 20}~=&~~ ~~^2{\bf 8}\,\oplus\, ^4{\bf 1}\,,
\end{align}
where the superscript $(2S+1)$ gives the spin for each particle in the
SU(3)~multiplet. The spin and parity $J^P =
\frac{1}{2}^+$~ground-state octet of
the {\bf 56}-plet, characterized by zero orbital angular momentum, 
consists of the proton, neutron, $\Lambda$, $\Sigma^{0,\pm}$, and
$\Xi^{0,-}$. The additional spin and parity $J^P =
\frac{3}{2}^+$~ground-state decuplet of
the {\bf 56}-plet, also corresponding to baryons with vanishing orbital angular momentum, 
contains the $\Delta^{0,\pm}$, $\Delta^{++}$, $\Sigma(1385)^{0,\pm}$, $\Xi(1530)^{0,-}$, and $\Omega^-$.
Both the octet and decuplet contain a $\Xi$~baryon, and for this
reason, the number of $\Xi$~resonances is expected to be equal to the
number of $N^\ast$ and $\Delta^\ast$~states combined.
Some excitation of the spatial part is required for the wave functions 
of the {\bf 70}-plet and {\bf 20}-plet to make the overall non-color (spin\,$\times$\,space\,$\times$\,flavor) component 
of the wave function symmetric. Orbital motion is incorporated by
classifying states into SU$(6)\,\otimes\,$O(3) supermultiplets, where the O(3)
group accounts for the orbital degrees of freedom. 

\begin{table}[b]
\caption{\label{chap3:tab1}Supermultiplets, $\Big(\,{\bf D},\,L_N^P\,\Big)$,  
  that occur in the first two excitation bands and assignments
  for the ground-state octet and decuplet to known baryons.}
\centering
{\addtolength{\extrarowheight}{4pt}
\begin{tabular}{@{}l|cccccc}
  \toprule
 N & \multicolumn{1}{c|}{SU(3)$_f$} & \multicolumn{5}{c}{Supermultiplets}\\[1ex]\hline
0 & \multicolumn{1}{c|}{} & \multicolumn{5}{c}{$\Big(\,{\bf 56},\,0_0^+\,\Big)$}\\[1ex]
   & \multicolumn{1}{c|}{$^2{\bf 8}$} & $S = \frac{1}{2}^+$ & $N(939)$ & $\Lambda(1116)$ & $\Sigma(1193)$ & $\Xi(1318)$\\[0.5ex]
   & \multicolumn{1}{c|}{$^4{\bf 10}$} & $S = \frac{3}{2}^+$ & $\Delta(1232)$ & $\Sigma(1385)$ & $\Xi(1530)$ & $\Omega(1672)$\\[1ex]\hline
  1 & & \multicolumn{5}{c}{$\Big(\,{\bf 70},\,1_1^- \,\Big)$}\\[1ex]\hline
2 & & $\Big(\,{\bf 56},\,0_2^+\,\Big)$ & $\Big(\,{\bf 70},\,0_2^+\,\Big)$ & $\Big(\,{\bf 20},\,1_2^+\,\Big)$ & $\Big(\,{\bf 70},\,2_2^+\,\Big)$ & $\Big(\,{\bf 56},\,2_2^+\,\Big)$\\[1ex]
\botrule
\end{tabular}}
\end{table}

Baryons can furthermore be organized into harmonic-oscillator
excitation bands, labeled by the number of quanta $N=0,1,2,...$ Each
band consists of a number of supermultiplets, specified by
$\big(\,{\bf D},\,L^P_N\,\big)$, where ${\bf D}$ is the dimensionality of the SU(6) spin-flavor representation, $L$ 
is the total quark orbital angular momentum, and $P$ is the parity. The first-excitation band contains only 
one supermultiplet, $\big(\,{\bf 70},\,1_1^-\,\big)$, corresponding to states with one unit of orbital angular momentum 
and negative parity, whereas the second-excitation band contains already five supermultiplets corresponding 
to states with positive parity and either two individual units of angular momentum that can couple to 
$L = 0, 1, 2$, giving the supermultiplets $\big(\,{\bf 70},\,0_2^+\,\big)$, $\big(\,{\bf 20},\,1_2^+\,\big)$, $\big(\,{\bf 70},\,2_2^+\,\big)$, 
respectively, two direct units of orbital angular momentum giving the supermultiplet $\big(\,{\bf 56},\,2_2^+\,\big)$, 
or one unit of radial excitation, $\big(\,{\bf
  56},\,0_2^+\,\big)$. Table~\ref{chap3:tab1} summarizes the
supermultiplets contained in the first three excitation bands,
together with the known, well-established baryons belonging to the
ground-state {\bf 56}-plet.

\subsubsection{Heavy-flavor baryons\label{BaryonClassification:HeavyFlavors}}
Baryons containing a single charm quark can still be organized into SU(3)$_f$
flavor multiplets. These multiplets, however, form only a subset of
the larger SU(4)$_f$ symmetry group, which encompasses baryons containing
zero, one, two, or three charm quarks. The same multiplet structure is
expected to occur for each combination of spin and parity, giving rise
to a rich spectrum of baryon states.
Analogous SU(4)$_f$ multiplets can be constructed by replacing the
charm quark with a bottom
quark, and both sets of SU(4)$_f$ representations may in turn be embedded
within a larger SU(5)$_f$ symmetry that includes all baryons composed of
the five quark flavors accessible at low and intermediate energies.
However, it is important to emphasize that the classification of baryons
into SU(4)$_f$ and SU(5)$_f$ multiplets is primarily of organizational value,
since these symmetries are strongly broken by the large mass
differences between the quarks. Only the approximate SU(3) $(u,d,s)$ and
SU(2) $(u,d\,)$ flavor symmetries are sufficiently well realized to
provide quantitative insight into the structure and properties of
baryons. The multiplet structure for flavor SU(4) is
\begin{eqnarray}
  {\bf 4}\,\otimes\,{\bf 4}\,\otimes {\bf 4} = {\bf
  20}_S\,\oplus\,{\bf 20}_M\,\oplus\,{\bf 20}_M\,\oplus\,{\bf 4}_A.
\end{eqnarray}
These decompose into irreducible flavor-SU(3) multiplets
\begin{align}
{\bf 20}_S\,=&~~ {\bf 10} \,\oplus\, {\bf 6}\,\oplus\, {\bf 3}\,\oplus\, {\bf 1}\\\label{Equation:20S-plet}
{\bf 20}_M\,=&~~ ~~{\bf 8} \,\oplus\, {\bf 6}\,\oplus\, {\bf 3}\,\oplus\, {\bf 3}\\\label{Equation:20M-plet}
{\bf 4}_A\,=&~~ ~~{\bf 3}\,\oplus\, {\bf 1}\,.
\end{align}

\begin{figure}[t]
\centerline{\includegraphics[width=0.78\textwidth]{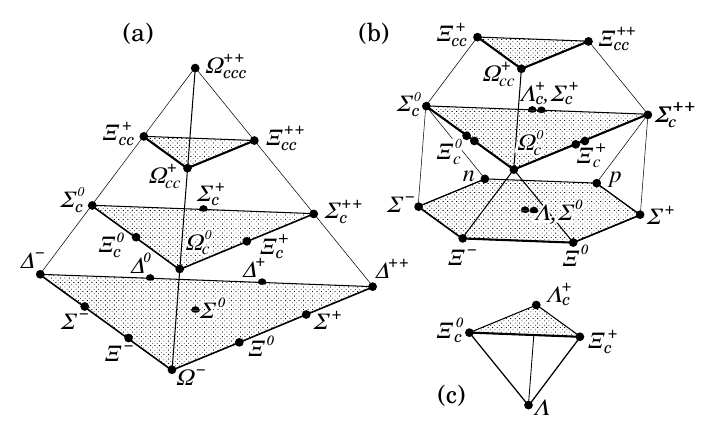}}
\caption{\label{chap3:fig1} Flavor-SU(4) multiplets: (a) The symmetric {\bf 20} of flavor SU(4), showing the SU(3)
  decuplet on the lowest layer. (b) The mixed-symmetric {\bf 20}s
  and (c) the antisymmetric {\bf 4} of SU(4). The mixed-symmetric {\bf
    20}s have the SU(3) octet on the lowest layer, while the {\bf 4} has
  the SU(3) singlet at the bottom. Note that there are two $\Xi_c^+$ and
  two $\Xi_c^0$~resonances on the middle layer of the mixed-symmetric
  {\bf 20}. Reproduced from [\cite{Crede:2013kia}]. \copyright~IOP Publishing Ltd.
  All rights reserved.}
\end{figure}

The SU(4)$_f$~multiplets are shown in Fig.~\ref{chap3:fig1}
in the conventional representation, with the charm quantum number $C$
along the vertical.
The symmetric {\bf 20}-plet in Figure~\ref{chap3:fig1}~(a) contains the flavor-SU(3) decuplet as a
subset, forming the {\it ground floor} of the weight diagram. All ground-state baryons
in this multiplet are associated with spin-parity
$J^P=\frac{3}{2}^+$. In addition to $\Delta$, $\Sigma^\ast$, $\Xi^\ast$, and
$\Omega$ $(C=0)$, this {\bf 20}-plet contains $\Sigma^\ast_c$,
$\Xi^\ast_c$, and $\Omega^\ast_c$ with charm content $C=1$; $\Xi_{cc}^\ast$
and $\Omega_{cc}^\ast$ with charm content $C=2$; and $\Omega_{ccc}^{++}$
at the top with charm content $C=3$. 
In contrast, the mixed-symmetric {\bf 20}-plet shown in
Figure~\ref{chap3:fig1}~(b) contains the flavor-SU(3) octet on
the lowest level of the weight diagram. All ground-state baryons in
this multiplet are associated with $J^P=\frac{1}{2}^+$. In addition to the
$N$, $\Lambda$, $\Sigma$, and $\Xi$~baryons (light-flavor ground-state octet), this {\bf 20}-plet contains 
$\Sigma_c$, $\Xi^{\,\prime}_c$, and $\Omega_c$ (sextet with charm content
$C=1$), $\Lambda_c,\Xi_{c}$ (triplet with charm content $C=1$), and
$\Xi_{cc},\Omega_{cc}$ (another triplet with charm content $C=2$). 
The fundamental {\bf 4}-plet with $J^P=\frac{1}{2}^-$ shown in
Fig.~\ref{chap3:fig1}~(c) decomposes into a light-quark
singlet and a triplet of charmed baryons.

In conclusion, we briefly note that the triply strange ground-state baryon $\Omega^-$ 
can be described within flavor SU(3) as a fully symmetric state. Consequently, SU(3)$_f$ predicts a single ground state, which
is indeed observed experimentally as $\Omega(1670)$. A closely
analogous situation is found for the
$I=\frac{3}{2}~\Delta$~resonances. In particular, the $|ddd\rangle$ state corresponding to the $\Delta^{-}$ 
(and similarly the $\Delta^{++} = |uuu\rangle$) played an important
role in motivating the introduction of the color degree of freedom
and, in turn, led to the prediction of the $\Omega^- = |sss\rangle$~state.
In contrast, the situation is somewhat different for singly heavy
baryons such as $\Omega_c = |ssc\rangle$ (and analogously $\Omega_b =
|ssb\rangle$). Experimentally, these systems exhibit two ground
states, namely $\Omega_c\,\frac{1}{2}^+$ and $\Omega_c(2770)\,\frac{3}{2}^+$.
This doubling of states can be naturally understood in the
heavy-quark, light-diquark picture, in which the two light quarks
form a spin-1 diquark subsystem that couples to the spin-$\frac{1}{2}$ 
heavy quark.

\subsubsection*{Singly heavy baryons containing a charm or bottom quark}
Baryons containing a single heavy quark (charm or bottom) are
particularly well suited for a description within potential models,
which provide fairly reliable predictions for their mass spectra. In
such approaches, the baryon can be effectively viewed as a system
consisting of an approximately static heavy quark and a comparatively
light, tightly bound quark-quark subsystem (a diquark). This
separation allows for treating the heavy quark, carrying well-defined
quantum numbers $J^P = \frac{1}{2}^+$, independently of the light degrees of freedom.
The two light quarks form a diquark with intrinsic spin $S=0$ or $S=1$,
corresponding to antisymmetric and symmetric spin wave functions,
respectively. The associated flavor structure of the light diquark can be organized according to
\begin{equation}
\label{Equation:diquarkStructure}
\mathbf{3} \,\otimes\, \mathbf{3} \,=\, \overline{\mathbf{3}}_A \,\oplus\, \mathbf{6}_S ,
\end{equation}
where the {\bf 6} representation is symmetric and the {$\overline{\mathbf{3}}$}
  representation is antisymmetric in flavor space. Furthermore, the color
  wave function of the diquark is totally antisymmetric. The inclusion
  of orbital angular momentum in the light diquark system modifies the total spin and parity of
  the light degrees of freedom. Finally, these light degrees of
  freedom are then combined with the heavy
  quark, giving rise to a {\it hyperfine} splitting in a singly heavy baryon.

\begin{figure}[b]
  \centering
  \begin{minipage}[b]{0.4\textwidth}
\begin{align*}
\vec{\rho}~=&~~\frac{1}{\sqrt{2}}\left(\vec{r}_1-\vec{r}_2\right),\qquad \mu_\rho \,=\, m_q
              \nonumber\\[3ex]
\vec{\lambda}~=&~~\frac{1}{\sqrt{6}}\left(\vec{r}_1+\vec{r}_2-2\vec{r}_3\right),
  \qquad \mu_\lambda \,=\, \frac{3m_q\,m_Q}{2m_q + m_Q}\\[-2.5ex]
\end{align*}
  \end{minipage}
  \begin{minipage}[b]{0.4\textwidth}
      \centering
      \includegraphics[width=0.44\textwidth]{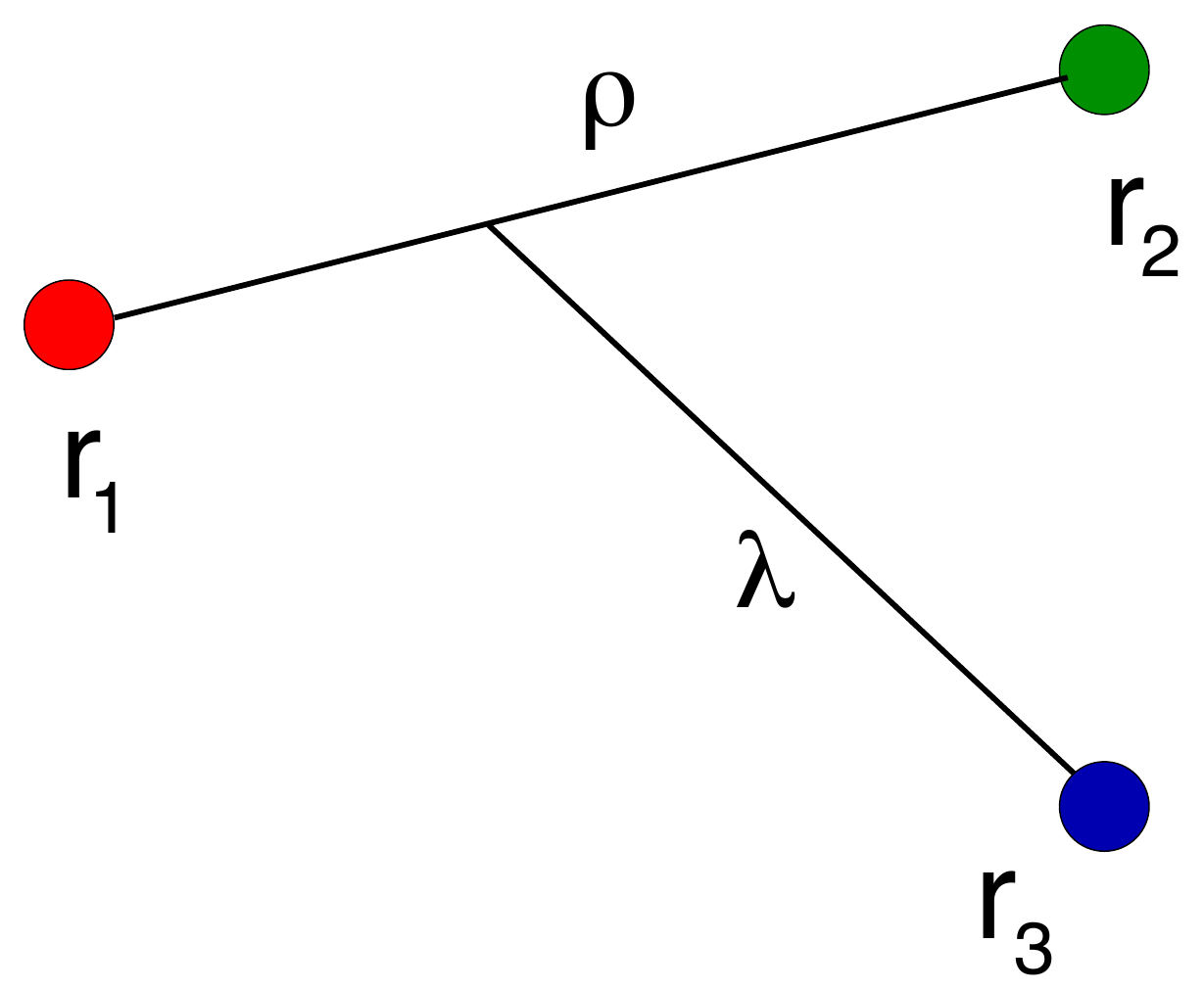}
      \vspace{0.1cm}
  \end{minipage}
  \caption{Simple quark-model depiction of a baryon, where the vectors
    $\vec{r}_i$ denote the positions of the three quarks, and
    $\mu_\rho$ and $\mu_\lambda$ denote the corresponding reduced masses. Here,
    $\vec{\rho}$ is proportional to the separation between quarks~1
    and 2, and $\lambda$ is proportional to the separation between
    quark~3 and the centre of mass of quarks~1 and 2. Orbital
    excitations in these non-relativistic systems can therefore be
    described in terms of the $\lambda$~mode and the
    $\rho$~mode.\label{chap3:fig2}}
\end{figure}

The non-relativistic dynamics of the quarks in the baryon can be described in terms of a
set of relative coordinates, $\vec{\rho}$ and 
$\vec{\lambda}$, so-called Jacobi coordinates, which are used to
simplify the description of multi-particle systems. They separate the
overall center-of-mass motion from the internal motion of the
particles. This simple picture of the baryon is depicted in
Fig.~\ref{chap3:fig2}, together with the definitions of $\vec{\rho}$ and 
$\vec{\lambda}$, and the corresponding reduced masses.
The total orbital angular momentum is then $\vec{L} = \vec{l}_\rho \otimes \vec{l}_\lambda$.
In the discussion of singly heavy baryons, the assignment of $\lambda$
and $\rho$ is straightforward:
The former corresponds to an excitation between the (static) heavy
quark and the light diquark system, whereas the latter describes the
internal excitation of the light diquark.
The mixing between the configurations appears to be small.

In the singly heavy baryon sector, it is indeed more useful to describe the
three-quark system in terms of a quark-diquark picture. The spectrum
can then be easily classified based on the flavor structure of the
light diquark. For the depiction of the baryon in Fig.~\ref{chap3:fig2},  $q =
u,d$ and $Q = c,b$ for the singly heavy charmed or bottom baryons, respectively. 
The ratio of the harmonic oscillator frequencies is given by
\begin{equation}
    \frac{\omega_\lambda}{\omega_\rho} \,=\, \sqrt{\frac{1}{3}\,(1+2m_q/m_Q})~\leq~1\,. 
\end{equation}
In the limit of $m_q\approx m_Q$, e.g. for $N^\ast$ and
$\Delta^\ast$~resonances, the excitation energies in the $\rho$ and
$\lambda$~oscillators are about the same and strongly mix, whereas the
excitation energies in the $\lambda$~oscillator are reduced by a
factor of $\sqrt{3}$ in the heavy-quark limit, $m_Q\to\infty$.

The light diquark obeys the Pauli Principle, and the structure of this
system is straightforward. The diquark has an antisymmetric
${\bf{\bar{3}}}_C$ color structure and either a symmetric
${\bf{6}}_F$ or an antisymmetric ${\bf{\bar{3}}}_F$ flavor structure,
see Eq.~(\ref{Equation:diquarkStructure}). Moreover, the spin
component of the diquark has either a symmetric $(s_{qq} = 1)$ or an
antisymmetric $(s_{qq} = 0)$ spin angular momentum structure. The
orbital angular momentum $l_\rho$ of the diquark can then be combined
with the spin angular momentum to yield either a scalar or an
axial-vector diquark. These diquarks are often considered in the
literature a {\it good} or a {\it bad}
diquark~[\cite{Gross:2022hyw,Chen:2016spr,Jaffe:2004ph}], respectively,
since a scalar diquark has an attractive interaction making the system
more tightly bound, whereas an axial-vector diquark has a repulsive
interaction. As an example, the $S$- and $P$-wave diquark structure is
given by:\\[-1.4ex]
\begin{eqnarray}
  l_\rho = 0~(S)=\left\{
    \begin{array}{@{} l c @{}}
      ~s_{qq} = 0~(A),~~{\bf{\bar{3}}}_F~(A) & j_{qq} = 0\,,\\[1ex]
      ~s_{qq} = 1~(S),~~{\bf{6}}_F~(S) & j_{qq} = 1\,,
    \end{array}\right.\qquad\qquad
  l_\rho = 1~(A)=\left\{
    \begin{array}{@{} l c @{}}
      ~s_{qq} = 0~(A),~~{\bf{6}}_F~(S) & j_{qq} = 1\,,\\[1ex]
      ~s_{qq} = 1~(S),~~{\bf{\bar{3}}}_F~(A) & j_{qq} = 0/1/2\,,
    \end{array}\right.
\end{eqnarray}
where $j_{qq} = s_{qq} \otimes l_\rho$ denotes the total angular
momentum of the diquark. Finally, the total angular momentum $J$ of
the singly heavy baryon is then
\begin{equation}
    J = s_Q \otimes (j_{qq} \otimes l_\lambda)\,.
\end{equation}
Tentative $\lambda$ and $\rho$ classifications of known charmed and bottom
baryons are given in Table~\ref{chap3:tab2}. In the bottom sector, the
$\Xi_b^{\,\prime\,0}\,\frac{1}{2}^+$ and $\Omega_b^\ast\,\frac{3}{2}^+$ ground states are still missing.

\begin{table}[t]
\caption{\label{chap3:tab2} Tentative $\lambda$ and
  $\rho$ classifications of charmed and bottom
  baryons, where $l_\rho,l_\lambda$ denote the orbital angular
  momentum of the two oscillators, $s_{qq}$ and $j_{qq}$ denote the
  spin and the total angular momentum of the diquark, respectively,
  and $L$ and $J^P$ are the total orbital angular momentum and the
  total spin of the baryon. The table has been adapted from
  Ref.~[\cite{Crede:2024hur}]. States labeled with $^\dagger$ are
  poorly established one- or two-star states. The PDG does not list
  $J^P$~quantum numbers for the color-highlighted states and the
  classifications are merely educated guesses.}
\centering
{\addtolength{\extrarowheight}{4pt}
\begin{tabular}{@{}c|ccccccc||c|c}
  \multicolumn{10}{c}{}\\[-2ex]
  \hline
 & $L$ & $l_\rho$ & $l_\lambda$ & $s_{qq}$ & $j_{qq}$ & $J^P$ & $(nL)$
 & Charmed Candidates & Bottom Candidates \\[1ex]
\hline
 & 0 & 0 & 0 & 0 & 0
 & $\frac{1}{2}^+$ & $(1S)$ & $\Lambda_c\,\frac{1}{2}^+$,~~$\bigl(\,\Xi_c^+,\Xi_c^0\,\bigr)\,\frac{1}{2}^+$
 & $\Lambda_b\,\frac{1}{2}^+$,~~$\bigl(\,\Xi_b^-,~\Xi_b^0\,\bigr)\,\frac{1}{2}^+$\\[1ex]
  & 0 & 0 & 0 & 0 & 0 & $\frac{1}{2}^+$ & $(2S)$
  & \bl{$\Lambda_c(2765)^\dagger$},~~$\Xi_c(2970)\,\frac{1}{2}^+$
  & $\Lambda_b(6070)^0\,\frac{1}{2}^+$\\[1ex]
  & & & & & & & & & \bl{$\Xi_b(6227)^-$},~~\bl{$\Xi_b(6227)^0$}\\[1ex]
  & 1 & 0 & 1 & 0 & 1 & $\frac{1}{2}^-$, $\frac{3}{2}^-$ & $(1P)$
  & $\Lambda_c(2595)^+\,\frac{1}{2}^-$,~~$\Lambda_c(2625)^+\,\frac{3}{2}^-$
  & $\Lambda_b(5912)^0\,\frac{1}{2}^-$,~~$\Lambda_b(5920)^0\,\frac{3}{2}^-$\\[1ex]
  & & & & & & & & $\Xi_c(2790)\,\frac{1}{2}^-$,~~$\Xi_c(2815)\,\frac{3}{2}^-$
  & $\Xi_b(6100)^-\,\frac{3}{2}^-$\\[1ex]
  \rb{$\bf{\bar{3}}_F$} & 1 & 1 & 0 & 1 & 0 & $\frac{1}{2}^-$ & $(1P)$ & &\\[1ex]
  $\Lambda_c^{(\ast)}$ & 1 & 1 & 0 & 1 & 1 & $\frac{1}{2}^-$, $\frac{3}{2}^-$ & $(1P)$ & &\\[1ex]
  $\Xi_c^{(\ast)}$ & 1 & 1 & 0 & 1 & 2 & $\frac{3}{2}^-$, $\frac{5}{2}^-$ & $(1P)$
  & \rbb{\hspace{0cm}$\biggr\}~~~\Lambda_c(2940)^+\,\frac{3}{2}^-$~~(possibly a $2P$ state)} &\\[3ex]
  & 2 & 0 & 2 & 0 & 2 & $\frac{3}{2}^+$, $\frac{5}{2}^+$ & $(1D)$
  & $\Lambda_c(2860)^+\,\frac{3}{2}^+$, $\Lambda_c(2880)^+\,\frac{5}{2}^+$
  & $\Lambda_b(6146)^0\,\frac{3}{2}^+$,~~$\Lambda_b(6152)^0\,\frac{5}{2}^+$\\[1ex]
  & & & & & & & & $\Xi_c(3055)\,\frac{3}{2}^+$,~~\bl{$\Xi_c(3080)$}
  & \bl{$\Xi_b(6327)^0$},~~\bl{$\Xi_b(6333)^0$}\\[1ex]
\hline
  & 0 & 0 & 0 & 1 & 1 & $\frac{1}{2}^+$, $\frac{3}{2}^+$ & $(1S)$
  & $\Sigma_c(2455)\,\frac{1}{2}^+$,~~$\Sigma_c(2520)\,\frac{3}{2}^+$
  & $\Sigma_b\,\frac{1}{2}^+$, $\Sigma_b^\ast\,\frac{3}{2}^+$\\[1ex]
  & & & & & & &
  & $(\,\Xi_c^{\,\prime\,+},\Xi_c^{\,\prime\,0}\,)\,\frac{1}{2}^+$,~~$\Xi_c(2645)\,\frac{3}{2}^+$
  & $\bigl(\,\Xi_b^{\,\prime}(5935)^-,\,?\,\bigr)\,\frac{1}{2}^+$\\[1ex]
  & & & & & & & & & $\bigl(\,\Xi_b(5945)^0,~\Xi_b(5955)^-\,\bigr)\,\frac{3}{2}^+$\\[1ex]
  \rb{$\bf{6}_F$} & & & & & & & & $\Omega_c^0\,\frac{1}{2}^+$, $\Omega_c(2770)^0\,\frac{3}{2}^+$
  & $\Omega_b^-\,\frac{1}{2}^+$, ?\\[1ex]
  $\Sigma_c^{(\ast)}$ & 1 & 0 & 1 & 1 & 1 & $\frac{1}{2}^-$, $\frac{3}{2}^-$ & $(1P)$ & ~\multirow{3}{*}{$\left(
    \begin{array}{@{} c @{}}
      ~~\bl{\Omega_c(3000,~3050)^0},~~\bl{\Xi_c(2882,~2923)^\dagger}~~\\[1ex]
      ~~\bl{\Omega_c(3065,~3090)^0}\\[1ex]
      ~~\bl{\Omega_c(3120,~3185)^0}
    \end{array}\right)^?$} & \multirow{2}{*}{
  $\left(
     \begin{array}{@{} c @{}}
       ~~\bl{\Omega_b(6316)^-},~~\bl{\Omega_b(6330)^-}~~\\[1ex]
       ~~\bl{\Omega_b(6340)^-},~~\bl{\Omega_b(6350)^-}~~
    \end{array}\right)^?$
  }\\[1ex]
  $\Xi_c^{\,\prime (\ast)}$ & 1 & 0 & 1 & 1 & 2 & $\frac{3}{2}^-$, $\frac{5}{2}^-$ & $(1P)$ & &\\[1ex]
  $\Omega_c^{(\ast)}$ & 1 & 1 & 0 & 0 & 1 & $\frac{1}{2}^-$, $\frac{3}{2}^-$ & $(1P)$ & &\\[2ex]
\hline
\end{tabular}}
\end{table}

\subsection{Experimental methods for studying baryons and baryon-like systems}
One of the earliest and historically most important methods for
studying baryons involved hadron-beam experiments. In such
experiments, pion, Kaon, or proton beams are directed onto fixed
(nuclear) targets, producing excited baryon states through the strong
interaction. Bubble chamber experiments in the 1950s and 1960s played
a crucial role in discovering many baryon resonances, including the
$\Omega^-$~baryon in the reaction $K^-\,p \to \Omega^-
\,K^+\,K^0$. Modern hadron facilities continue this approach with more
advanced experimental setups using
sophisticated magnetic spectrometers and particle detectors capable of
reconstructing complex final states with high precision.

Light-flavor baryons are predominantly investigated in fixed-target
experiments employing hadronic or electromagnetic probes, whereas the
production of heavy-flavor baryons requires substantially higher
energies and is therefore primarily carried out at collider
facilities. However, new opportunities for studying light-flavor,
multi-strange hyperons have also emerged in recent years from the high-energy
experiments such as Belle, BES\,III, and LHCb. At these facilities,
light-flavor baryon resonances are produced abundantly in the decays
of heavy baryons containing charm or bottom quarks, as well as in
decays of heavy-flavor mesons. The large, high-statistics data samples
collected by these experiments provide an exceptionally rich
environment for investigating the properties of
multi-strange hyperons.

\subparagraph{1. Fixed-target experiments}
Light-flavor baryons can be investigated through pion-induced
reactions from the nucleon at dedicated hadron facilities (active as of 2026)
such as the HADES experiment at the Center for Heavy Ion Research
(GSI) in Germany and experiments at the Japan Proton Accelerator Research Complex
(J-PARC). Future hadron facilities include experiments employing
$K_L$~beams at Jefferson Lab, $K^-$~beams at J-PARC, and proton beams at GSI.
Alternatively, electron scattering experiments constitute
another powerful method for probing baryons. Because electrons
interact electromagnetically rather than strongly, they provide a
comparatively clean probe for baryon structure studies. In elastic
electron–proton scattering, measurements of electromagnetic form
factors reveal information about the spatial distributions of charge
and magnetization inside the nucleon. Inelastic scattering experiments
allow the excitation of nucleon resonances, making it possible to
investigate the transition amplitudes between ground and excited
baryon states. Facilities such as Jefferson Laboratory have played a central role in studying nucleon
resonances through meson electroproduction processes.

Closely related are photoproduction experiments, in which
medium-energy photons are used to excite baryons in reactions such as
$\gamma N\to N^\ast/\Delta^\ast$ or in associated strangeness
production such as $\gamma N\to K\,Y^\ast$ (with
$Y^\ast = \Lambda^\ast, \Sigma^\ast$). In the latter reaction, the
hyperon can be produced at low energies in the decay of an intermediate
resonance ($s$-channel production) or at moderate energies without the
formation of an intermediate resonance ($t$-channel exchange processes)\,\footnote{In baryon-resonance
  production, the quantities $s$, $t$, and also $u$ refer to the
  so-called Mandelstam variables, which describe the kinematics of a
  scattering reaction in a relativistically invariant
  way. Qualitatively, in $s$-channel production, the incoming
  particles form an intermediate resonance, and in $t$-channel
  production, the incoming particles interact by exchanging another particle.}.
Tagged photon beams
incident on proton or nuclear targets produce excited nucleon and
hyperon states that subsequently decay into mesons and lighter
baryons. Since photons couple directly to electric charge,
photoproduction experiments are highly sensitive to the internal quark
dynamics of baryons. Additional information is provided by the
measurement of polarization observables in experiments that
use polarized beams and targets, for example, at Jefferson
Lab~[\cite{McKeown:2013sxa}], United States, the 
Electron Stretcher Accelerator (ELSA) in Bonn~[\cite{Hillert:2006yb}], Germany, the
Mainz Mikrotron (MAMI) in Mainz~[\cite{Mecking:2006yh}], Germany, the Grenoble Anneau Accelerateur
Laser (GRAAL) facility~[\cite{GRAAL:2005mor}], France, and at the Laser Electron Photon Experiment
hosted at SPring-8 (LEPS)~[\cite{LEPS:2013tbd}], Japan. These observables play a crucial role 
in the search for new baryon resonances and in the determination of their spin and
parity quantum numbers. The extraction of resonance properties requires
sophisticated
partial-wave analyses that are capable of 
disentangling contributions of several broad and overlapping baryon states.
These analyses combine experimental data from
multiple reaction channels with theoretical constraints derived from
unitarity, analyticity, and symmetry principles to determine their masses, widths, spin-parity assignments, and
coupling strengths. 

\subparagraph{2. Collider experiments}
The study of heavy-flavor baryons containing charm or bottom quarks has
become an increasingly important area of experimental research in
hadron spectroscopy. In addition to a plethora of new conventional
baryons, the observation of several charmed pentaquark candidates, and
even a doubly
charmed baryon, at the Large
Hadron Collider is particularly exciting.
Heavy baryons are produced in
high-energy proton-proton collisions but also in electron-positron annihilation
experiments. Worth noting again are the opportunities for studying
multi-strange hyperons in the decay of charmed baryons in reactions such
as $\Xi_c^+\,(\Lambda_c^+)\to [\Xi^- \pi^+]_{\,\Xi^\ast}\,\pi^+ \,(K^+)$
or the observation of excited nucleon states in the decay of charmonia, $e^+e^- \to
J/\psi \to N^\ast \bar{N}$.

Bottom baryons are too heavy to be produced at the
$B$~factories that are optimized to produce large, clean samples of
$B$~mesons for precision studies of heavy-flavor physics, while
higher-energy $e^+ e^-$~colliders such as the
Large Electron–Positron (LEP) collider at CERN suffered from
relatively low luminosities in the past. As a result, studies of $B$~baryons have
primarily been carried out at hadron colliders. Experiments at
Fermilab provided some of the earliest investigations of bottom
baryons, although these studies were largely secondary to their main
physics programs.
The highest energies are now available at the LHC. Here, the
general-purpose experiments ATLAS, CMS, and LHCb are well positioned
to make significant contributions to our understanding of heavy-hadron
production and decays. In particular, the LHCb experiment is optimized for
reconstructing the decay vertices of heavy-flavor hadrons with
extremely high precision. 

\subsection{Experimental status of baryon systems}
Baryon resonances are classified by the Particle Data Group using a
``star'' rating system to indicate how confidently a particle's
existence and properties have been confirmed
experimentally. Consequently,
$\ast\ast\ast\,\ast$~baryons are considered well established, with strong and
consistent evidence from multiple experiments ({\it ``existence is
  certain''}~[\cite{ParticleDataGroup:2024cfk}]), while $\ast\ast\,\ast$~states
are fairly well established but may still have some uncertainties in
their measured quantum numbers ({\it ``existence ranges from very likely to
certain, but further confirmation is desirable and\,/\,or quantum numbers,
branching fractions, etc. are not well
determined''}~[PDG]). Furthermore, $\ast\,\ast$~baryons are based on only fair
evidence, often relying on limited or indirect experimental data ({\it
``evidence of existence is only fair''}~[PDG]), and $\ast$~states are poorly established candidates that require further
confirmation ({\it``evidence of existence is poor''}~[PDG]). These star assignments help physicists evaluate the
reliability of reported resonances in the complex baryon spectrum,
where many excited states overlap and decay rapidly through the strong
interaction. With the exception of low-mass ground states, peaking
structures in mass distributions associated with baryons usually
consist of several contributing resonances. For this reason, certain
mass regions in light-flavor baryon spectroscopy are merely referred to as {\it resonance regions},
e.g., beyond the $\Delta$~mass region, the mass regions close to
1500~MeV, 1700~MeV, and 1900~MeV are called the {\it first resonance
  region}, {\it second}, and {\it third}, respectively.

In the 1960s and 1970s, deep inelastic high-energy neutrino and electron scattering
experiments on the nucleon
were instrumental in establishing the existence of quarks. However, a detailed
understanding of the quark dynamics within nucleons required both more
advanced theoretical concepts and accelerator technology. Modern
nucleon-structure studies are now carried out at facilities such as
Jefferson Lab, where the high-quality CEBAF beam enables precision
investigations of deeply virtual meson production and the extraction
of nucleon structure functions. A comprehensive review of the nucleon
spin structure can be found in Ref.~[\cite{Deur:2018roz}].
Unfortunately, the collective
degrees of freedom of the nucleon are largely inaccessible in deep
inelastic scattering experiments. Thus, baryon spectroscopy provides
complementary insight into the existence and properties of excited
baryon states. Current experimental programs focus on mapping the
excitation spectrum of both light- and heavy-flavor baryons, and on
identifying possible underlying symmetries of the spectrum. In recent
years, extensive new data on $N^\ast$ and $\Delta^\ast$~resonances have
been collected at facilities worldwide, including Jefferson
Lab, ELSA, MAMI, GRAAL, and LEPS. A central objective of the
global $N^\ast$~program has been the realization of so-called {\it complete
experiments,} which allow for the unambiguous extraction of scattering
amplitudes and resonance properties in meson photo- and
electroproduction reactions.

In baryon spectroscopy, the concept of a
complete experiment arises from the need to determine the underlying
reaction amplitudes in
a model-independent way. In reactions such as $\gamma N \rightarrow
\pi N$, $\eta N$, $\omega N$, $\pi\pi N$, $K\Lambda$, or $K\Sigma$, the scattering process is
described by four complex helicity or CGLN amplitudes\,\footnote{The
  Chew–Goldberger–Low–Nambu (CGLN) amplitudes of
  Ref.~[\cite{Chew:1957tf}] provide a standard decomposition of the
  pseudoscalar meson photoproduction amplitude in terms of four
  complex scalar functions and are usually denoted by $F_{1-4}$. They encode
  the full spin structure of studied reactions and are equivalent to
  other representations, such as helicity or transversity
  amplitudes.}, corresponding to eight real quantities at each energy
and scattering angle. Since an overall phase cannot be measured
experimentally, the measurement of at least eight carefully chosen
observables is
required to fully reconstruct the amplitudes, up to some remaining discrete
ambiguities~[\cite{Chiang:1996em}]. Of course, the measurement of only these eight
observables would merely address the mathematical aspect of the
problem but would neglect the experimental uncertainties in the data, which
in practice impose significant limitations on the precision and
stability of the extracted amplitudes.
Measurements of differential cross sections alone are
therefore insufficient, as they provide access only to bilinear
combinations of amplitudes and do not uniquely constrain the resonance
content of the reaction. Complete experiments consequently rely on
extensive measurements of single- and double-polarization observables,
obtained through combinations of polarized photon or electron beams,
polarized nucleon targets, and the detection of the polarization of
the recoiling baryon. Typical
observables include the beam asymmetry $\Sigma$, target asymmetry
$T$, recoil polarizations $P$, and various correlated beam-target or
beam-recoil observables.

Partial-wave analyses represent one of the most powerful tools for
extracting resonance contributions from experimental data. They
form the cornerstone of hadron spectroscopy and provide the
essential connection between experimental observables and the
underlying spectrum of excited states. Since baryon resonances
are unstable objects with large decay widths and observed baryon
structures are often based on contributions from many
overlapping states, baryons cannot generally be identified
directly from peaks in measured cross sections. Instead, the reaction
amplitudes must be decomposed into partial waves characterized by
definite quantum numbers, such as total angular momentum~$J$,
parity~$P$, and isospin~$I$. Partial-wave analyses are therefore indispensable for
extracting resonance parameters, including pole positions, decay
widths, branching ratios, and electromagnetic transition amplitudes,
from the large body of experimental data obtained in hadron-induced and
electromagnetic reactions. In practice, modern partial-wave analyses involve global
fits to extensive data sets comprising differential cross sections,
spin-density matrix elements, and numerous single- and
double-polarization observables. Contemporary analyses are commonly
based on coupled-channel approaches, in which several reaction
channels are analyzed simultaneously in order to account consistently
for channel couplings, rescattering effects, and threshold
phenomena. These approaches incorporate fundamental theoretical
principles such as unitarity, analyticity, gauge invariance, and
crossing symmetry, thereby ensuring a properly defined
description of the reaction dynamics.

In a brief summary, recent
measurements in photo- and electroproduction have significantly
extended and complemented earlier results in baryon spectroscopy
obtained from $\pi$- and $K$-induced reactions, and have improved the
identification and classification of excited states, particularly of
excited nucleon states. They
continue to provide stringent constraints for partial-wave analyses,
enabling the extraction of resonance properties and electromagnetic transition form factors with
significantly reduced model dependence. The comprehensive database of 
polarization observables obtained in recent years has therefore played a
crucial role in establishing new excited nucleon states and, more
importantly, in
refining the star assignments of known resonances and in improving our
understanding of the non-perturbative dynamics of quantum
chromodynamics in the confinement regime.

\subsubsection{Light-flavor, non-strange baryons}
\begin{table}
\begin{center}
\caption{\label{chap3:tab3}Tentative assignments of the known light baryons to the lowest-lying
  SU(6)\,$\otimes$\,O(3) octets, decuplets, and singlets. The table has been adapted from
  Ref.~[\cite{Gross:2022hyw}]. States labeled with $^\dagger$ are poorly established one- or two-star states.
  With the exception of the lowest-lying states, all multiplet assignments are merely educated guesses because
  either the evidence for their existence is poor or they can be assigned to other multiplets. A hyphen indicates
  that the state does not exist; an empty field means that it is missing. In the $\Omega$~sector, only the ground state
  has known quantum numbers; therefore, the $\Omega$~column has been omitted. Based on the discussion
  in Ref.~[\cite{Gross:2022hyw}], two pairs of
  $(1/2^{\,-},\,3/2^{\,-})$~$\Sigma$~states and a $(1/2^{\,-},\,3/2^{\,-},5/2^{\,-})$~triplet
  with masses below
  2100~MeV are expected, but only weak evidence for two pairs is observed. These states are labeled with $^{\,a}$ and
  two possible multiplet assignments are shown. Likewise, a pair of $(5/2^{\,-},\,7/2^{\,-})$ $N^\ast$~states
  cannot unambiguously be assigned to $\big(\,70,\,3_3^-\,\big)$~multiplets. These states are labeled with
  $^{\,b}$. The color-highlighted states have been proposed based on the recent photoproduction data.}
{\addtolength{\extrarowheight}{3pt}
\begin{tabular}{@{}r|l|c|c|c|c|c|c||c|c|c||c@{}}
  \multicolumn{12}{c}{}\\[-3ex]
\hline
$\,N$ & $(D,\,L^P_N)$ & $S$ & $J^P$ & \multicolumn{4}{c||}{Octet Members}
        & \multicolumn{3}{c||}{Decuplet Members} & Singlets\\[0.5ex]
  \hline
0 & $\big(\,56,\,0_0^+\,\big)$ & $1/2$ & $1/2^{\,+}$ & $N(939)$ & $\Lambda(1116)$
   & $\Sigma(1193)$ & $\Xi(1318)$ & $-$ & $-$ & $-$ & $-$\\
   & & $3/2$ & $3/2^{\,+}$ & $-$ & $-$ & $-$ & $-$ & $\Delta(1232)$ & $\Sigma(1385)$
   & $\Xi(1530)$ & $-$\\[0.5ex]
  \hline
1 & $\big(\,70,\,1_1^-\,\big)$ & $1/2$ & $1/2^{\,-}$ & $N(1535)$ & $\Lambda(1670)$
   & $\Sigma(1620)^\dagger$ & $\Xi(1690)$& $\Delta(1620)$ & $\Sigma(1900)^{\dagger\,a}$ &
   & $\Lambda(1405)$\\
   & & & $3/2^{\,-}$ & $N(1520)$ & $\Lambda(1690)$ & $\Sigma(1670)$ & $\Xi(1820)$
   & $\Delta(1700)$ & $\Sigma(1910)^{\,a}$ & & $\Lambda(1520)$\\
   & & $3/2$ & $1/2^{\,-}$ & $N(1650)$ & $\Lambda(1800)$ & $\Sigma(1750)$ & & $-$ & $-$ & $-$ & $-$\\
   & & & $3/2^{\,-}$ & $N(1700)$ & & & & $-$ & $-$ & $-$ & $-$\\
   & & & $5/2^{\,-}$ & $N(1675)$ & $\Lambda(1830)$ & $\Sigma(1775)$ & & $-$ & $-$ & $-$ & $-$\\[0.5ex]
  \hline
2 & $\big(\,56,\,0_2^+\,\big)$ & $1/2$ & $1/2^{\,+}$ & $N(1440)$ & $\Lambda(1600)$
   & $\Sigma(1660)$ & & $-$ & $-$ & $-$ & $-$\\
   & & $3/2$ & $3/2^{\,+}$ & $-$ & $-$ & $-$ & $-$ & $\Delta(1600)$ & $\Sigma(1780)^\dagger$ & & $-$\\
   & $\big(\,70,\,0_2^+\,\big)$ & $1/2$ & $1/2^{\,+}$ & $N(1710)$ & $\Lambda(1810)^\dagger$
   & $\Sigma(1880)^\dagger$ & & $\Delta(1750)^\dagger$ & & & $\Lambda(1710)^\dagger$\\
   & & $3/2$ & $3/2^{\,+}$ &  & & & & $-$ & $-$ & $-$ & $-$\\
   & $\big(\,56,\,2_2^+\,\big)$ & $1/2$ & $3/2^{\,+}$ & $N(1720)$ & $\Lambda(1890)$
   & $\Sigma(1940)^\dagger$ & & $-$ & $-$ & $-$ & $-$\\
   & & & $5/2^{\,+}$ & $N(1680)$ & $\Lambda(1820)$ & $\Sigma(1915)$ & & $-$ & $-$ & $-$ & $-$\\
   & & $3/2$ & $1/2^{\,+}$ & $-$ & $-$ & $-$ & $-$ & $\Delta(1910)$ & & & $-$\\
   & & & $3/2^{\,+}$ & $-$ & $-$ & $-$ & $-$ & $\Delta(1920)$ & $\Sigma(2080)^\dagger$ & & $-$\\
   & & & $5/2^{\,+}$ & $-$ & $-$ & $-$ & $-$ & $\Delta(1905)$ & $\Sigma(2070)^\dagger$ & & $-$\\
   & & & $7/2^{\,+}$ & $-$ & $-$ & $-$ & $-$ & $\Delta(1950)$ & $\Sigma(2030)$ & & $-$\\
   & $\big(\,70,\,2_2^+\,\big)$ & $1/2$ & $3/2^{\,+}$ &  & & & & & & & $\Lambda(2070)^\dagger$\\
   & & & $5/2^{\,+}$ & \bl{$N(1860)^\dagger$} & & & & $\Delta(2000)^\dagger$ & & & $\Lambda(2110)$\\
   & & $3/2$ & $1/2^{\,+}$ & \bl{$N(1880)$} & & & & $-$ & $-$ & $-$ & $-$\\
   & & & $3/2^{\,+}$ & $N(1900)$ & & & & $-$ & $-$ & $-$ & $-$\\
   & & & $5/2^{\,+}$ & $N(2000)^\dagger$ & & & & $-$ & $-$ & $-$ & $-$\\
   & & & $7/2^{\,+}$ & $N(1990)^\dagger$ & $\Lambda(2085)^\dagger$ & & & $-$ & $-$ & $-$ & $-$\\
   & $\big(\,20,\,1_2^+\,\big)$ & $1/2$ & $1/2^{\,+}$ & & & & & $-$ & $-$ & $-$ &\\
   & & $(3/2)$ & $3/2^{\,+}$ & & & & & $-$ & $-$ & $-$ &\\
   & & & $5/2^{\,+}$ & $-$ & $-$ & $-$ & $-$ & $-$ & $-$ & $-$ &\\[0.5ex]
  \hline
3 & $\big(\,56,\,1_3^-\,\big)$ & $1/2$ & $1/2^{\,-}$ & \bl{$N(1895)$}
   & $\Lambda(2000)^\dagger$ & $\Sigma(1900)^{\dagger\,a}$ & & $-$ & $-$ & $-$ & $-$\\
   & & & $3/2^{\,-}$ & \bl{$N(1875)$} & $\Lambda(2050)^\dagger$ & $\Sigma(1910)^{\,a}$
   & & $-$ & $-$ & $-$ & $-$\\
   & & $3/2$ & $1/2^{\,-}$ & $-$ & $-$ & $-$ & $-$ & $\Delta(1900)$ & $\Sigma(2110)^{\dagger\,a}$ & & $-$\\
   & & & $3/2^{\,-}$ & $-$ & $-$ & $-$ & $-$ & $\Delta(1940)^\dagger$ & $\Sigma(2010)^{\dagger\,a}$ & & $-$\\
   & & & $5/2^{\,-}$ & $-$ & $-$ & $-$ & $-$ & $\Delta(1930)$ & & & $-$\\
   & $\big(\,70,\,3_3^-\,\big)$ & $1/2$ & $5/2^{\,-}$ & \bl{$N(2060)^{\,b}$} & & & & & & & $\Lambda(2080)^\dagger$\\
   & & & $7/2^{\,-}$ & $N(2190)^{\,b}$ & & $\Sigma(2100)^\dagger$ & & $\Delta(2200)$ & & & $\Lambda(2100)$\\
   & & $3/2$ & $3/2^{\,-}$ & \bl{$N(2120)$} & & & & $-$ & $-$ & $-$ & $-$\\
   & & & $5/2^{\,-}$ & \bl{$N(2060)^{\,b}$} & & & & $-$ & $-$ & $-$ & $-$\\
   & & & $7/2^{\,-}$ & $N(2190)^{\,b}$ & & & & $-$ & $-$ & $-$ & $-$\\
   & & & $9/2^{\,-}$ & $N(2250)$ & & & & $-$ & $-$ & $-$ & $-$\\\cline{2-12}
   & \multicolumn{11}{c}{additional multiplets: $\big(\,70,\,1_3^-\,\big)$, $\big(\,70,\,1_3^-\,\big)$,
     $\big(\,20,\,1_3^-\,\big)$, $\big(\,70,\,2_3^-\,\big)$, $\big(\,56,\,3_3^-\,\big)$,
     $\big(\,20,\,3_3^-\,\big)$}\\[0.5ex]
  \hline
4 & & & & $N(2220)$ & $\Lambda(2350)$ & & & $\Delta(2420)$ & & &\\[0.5ex]
  \hline
5 & & & & $N(2600)$ & & & & $\Delta(2750)^\dagger$ & & &\\[0.5ex]
  \hline
\end{tabular}}
\end{center}
\end{table}

Mapping the spectrum of excited nucleon states and understanding the
effective degrees of freedom are very challenging tasks in
hadron physics. Elastic pion scattering, $\pi N \to \pi N$, provided the primary
experimental evidence for nearly all nucleon resonances included in
the Review of Particle Physics prior to the 2012 edition. Despite its historical
importance, this channel has limited sensitivity to higher-mass
excited nucleon states, particularly those with very small $\Gamma_{\pi
  N}$~decay widths,
whose signals become extremely difficult to isolate in elastic
scattering processes. Detailed partial-wave analyses performed by
just a small number of independent groups extracted the real and imaginary components
of the $\pi N$~amplitudes from available data, and these results
continue to serve as essential constraints in current studies of
photoproduction reactions. A detailed discussion of these partial-wave analyses, how resonances are
extracted from the data, and a review of individual resonances goes beyond the scope
of this chapter of the Encyclopedia of Nuclear Physics. Therefore, simply the 2026 status of
light-flavor baryon spectroscopy is presented here. For further reviews of the
field, we refer to the literature; see, for example, Ref.~[\cite{Gross:2022hyw}].

Table~\ref{chap3:tab3} summarizes the observed light-flavor baryon
resonances within the traditional SU(6)\,$\otimes$\,O(3) classification
scheme, which is based on the picture of non-relativistic constituent
quarks. The new additions based on recent photoproduction and
polarization data are color-highlighted. One of the enduring successes of the quark model is that this
framework describes the observed baryon spectrum remarkably well,
despite its conceptual simplicity. Even in relativistic approaches,
many baryon resonances retain dominant wave-function components that
can still be associated with SU(6)\,$\otimes$\,O(3) multiplets.
Also, lattice-QCD results exhibit the broad features expected of the
SU(6)\,$\otimes$\,O(3)~symmetry. The simple counting of excited states
appears consistent with non-relativistic quark
models~[\cite{Edwards:2012fx}].

The first excitation shell with negative parity, corresponding to
$N=1$, appears to be relatively complete, although several open
questions remain for the multi-strange hyperons. In agreement with
expectations, five $N^\ast$ and two
$\Delta^\ast$~negative-parity states are observed. Among the
negative-parity $\Lambda$ and $\Sigma$ octet states, only one $J^P =3/2^-$~pair 
remains unobserved. The two resonances $\Lambda(1800)\,1/2^-$ 
and $\Sigma(1750)\,1/2^-$ are commonly interpreted as states with
intrinsic quark spin $3/2$, forming spin partners of $\Lambda(1830)\,5/2^-$ and $\Sigma(1775)\,5/2^-$.
The identification of the negative-parity decuplet $\Sigma$~states is
less certain. Quark-model expectations suggest one decuplet doublet near
1750~MeV, together with an additional octet doublet and decuplet triplet around
2050~MeV in the $\big(\,56,\,1_3^-\,\big)$~configuration. However, current
analyses provide only limited evidence for such states, denoted by
the superscript~$^{\,a}$ in Table~\ref{chap3:tab3}. The singlet states
$\Lambda(1405)\,1/2^-$ and $\Lambda(1520)\,3/2^-$
are particularly intriguing and warrant special attention; see the next
section on hyperons.
At higher masses, the assignment of resonances becomes increasingly
ambiguous. For example, the $N(1900)\,3/2^+$~resonance is usually
associated with the second excitation shell, although it may
correspond to configurations with intrinsic quark spin $1/2$ or $3/2$,
both having orbital angular momentum $L=2$. In addition, a radially
excited $3/2^+$~state with $L=0$ is also expected in this mass region,
allowing for possible mixing among these states. Experimentally,
however, only one resonance has been clearly identified, despite quark
models predicting three $J^P =3/2^+$~states. Also absent is the
expected $L=1$ doublet belonging to the SU(6)\,$\otimes$\,O(3)
{\bf 20}-plet, whose production is believed to be strongly suppressed. Only
a limited number of hyperons can presently be assigned to
the second excitation shell, and although some $\Lambda$~states are
plausibly interpreted as SU(3) singlets, these assignments remain
tentative.

\paragraph{Do parity doublets exist?}
The appearance of parity doublets in the high-mass baryon spectrum
presents another important puzzle. In low-energy hadron physics,
spontaneous chiral symmetry breaking generates large mass splittings
between chiral partners, such as the $\rho(770)$ and
$a_1(1260)$~mesons or the nucleon and $N(1535)\,1/2^-$.
Surprisingly, many highly excited baryons appear in nearly degenerate
pairs with identical spin but opposite parity. Similar patterns are
also observed in the meson spectrum, motivating the hypothesis that
chiral symmetry may become effectively restored in highly excited
hadrons. If correct, this would imply that every sufficiently excited
resonance should possess a corresponding parity partner, providing a
clear and testable prediction; see, for example, Ref.~[\cite{Glozman:2003bt}].

\paragraph{How can three-quark resonances and dynamically generated
  resonances be reconciled?}
A further open question concerns the possible existence of dynamically
generated resonances\,\footnote{Dynamically generated baryon resonances are
  states that emerge primarily from strong meson–baryon degrees of freedom
  and coupled-channel dynamics rather than from conventional
  three-quark excitations predicted by constituent quark
  models.}. Beyond the well-known $\Lambda(1405)\,1/2^-$,
one of the earliest candidates was the negative-parity
$N(1535)\,1/2^-$~[\cite{Kaiser:1995cy}]. Its interpretation sparked considerable debate,
particularly between advocates of dynamical coupled-channel approaches
and supporters of the traditional quark model description. The idea
that there could be two different and overlapping states is not
supported by data. While some
studies suggest that both $N(1535)\,1/2^-$ and $N(1650)\,1/2^-$ 
could emerge dynamically from meson-baryon interactions~[\cite{Bruns:2010sv}], other
resonances such as $\Delta(1620)\,1/2^-$ do not appear to fit
naturally into this picture. This raises a fundamental question in
baryon spectroscopy: Are conventional three-quark resonance poles and
dynamically generated poles simply different descriptions of the same
physical states, or do they correspond to genuinely distinct and
possibly orthogonal configurations?

\subsubsection{Light-flavor baryons with non-zero strangeness (hyperons)}
Light-flavor baryons that contain at least one strange (valence) quark in
addition to up and\,/\,or down quarks are called hyperons and denoted
$\Lambda$, $\Sigma$ ($S = -1$), $\Xi$ ($S = -2$), and
$\Omega$ ($S = -3$). Many hyperons decay through the weak
interaction, typically with relatively long lifetimes of the order of
$10^{-10}$~s. Hyperons are more difficult to produce than non-strange
$N^\ast$ or $\Delta^\ast$~resonances because their production requires the creation of one or more
strange quarks. In processes that proceed through the strong or
electromagnetic interaction, strangeness is
conserved, so strange quarks cannot simply appear individually. They
must be produced together with corresponding anti-strange quarks, for example,
in the photoproduction reaction $\gamma p\,|duu\rangle\to K^+\,|\bar{s}u\rangle\,
\Lambda\,|sdu\rangle$ (``associated strangeness production''). This requires additional energy compared to
the production of ordinary non-strange hadrons composed only of up and
down quarks. As a result, reactions with hyperons in the final state have higher kinematic
thresholds and generally also much smaller production cross sections. A
more effective way is to start with a probe that already contains
strangeness such as a $K^-$~meson. However, dedicated high-intensity
$K^-$ beam facilities are rare today compared to pion- or proton-beam
facilities, and consequently only limited progress has been made in
the search for new hyperon states. On the experimental side, the study
of hyperons offers two distinct advantages compared to their
non-strange counterparts. Because hyperons decay weakly, they
typically have longer lifetimes than strongly decaying particles,
and therefore, they travel measurable distances before decaying inside the
detector. Although the reconstruction of their decay products requires
precise tracking capabilities, the presence of a detached decay vertex
provides a powerful tool for suppressing background
contributions. Furthermore, their weak decays violate parity, leading
to strongly asymmetric decay angular distributions. This property makes many
hyperons ``self-analyzing,'' in the sense that their polarization can be
extracted directly from the angular distributions of their decay
products without the need for a physical detector. Thus, hyperon production is especially valuable
for polarization studies in baryon spectroscopy despite the
experimental challenges.

\paragraph{Singly strange $\Lambda$ and $\Sigma$ resonances}
The study of hyperons complements studies of the nucleon excitation
spectrum. However, little progress has been reported in the hyperon
sector and the established $\Lambda$ and $\Sigma$~states reported by
the PDG in the latest edition of the
RPP~[\cite{ParticleDataGroup:2024cfk}] are essentially the same as
those that were listed
in the 1984 edition~[\cite{ParticleDataGroup:1984mfx}]. Despite the lack of new states, some
star ratings have been updated based on the results of recent
partial-wave analyses of older data. In the 2026 edition of the RPP, a
total of 14~$\Lambda$ and nine~$\Sigma$~states is
listed with an overall three-star or four-star ranking. In addition,
seven one-star and two two-star $\Lambda$~resonances, as well as 14
one-star and three two-star $\Sigma$~resonances, are known (omitted
from the Summary Table, though). We refer to
the review in Ref.~[\cite{Klempt:2020bdu}] for more details on the properties
of singly strange hyperons and multiplet assignments. The known
$\Lambda$ and $\Sigma$~baryons are shown in Table~\ref{chap3:tab3},
including tentative assignments to the lowest-lying
SU(6)\,$\otimes$\,O(3) octets, decuplets, and singlets.

For this chapter of the Encyclopedia of Nuclear Physics, the $\Lambda(1405)$ hyperon
merits further discussion since the resonance
occupies a unique and particularly intriguing position in baryon
spectroscopy. Discovered already in the early 1960s, it is an excited
$\Lambda$~state with spin-parity quantum numbers
$J^P=\frac{1}{2}^-$~[\cite{CLAS:2014tbc}], located just below the
$\bar{K}N$ threshold. Unlike most excited baryons, whose properties
can be reasonably well described within conventional quark models as
three-quark excitations, the $\Lambda(1405)$ has resisted a simple
interpretation for the longest time. Its unusually low mass, distorted
line shape in various isospin-related decay modes, and strong coupling
to the $\pi\Sigma$ channel suggest that it may not be an ordinary
three-quark state but rather a dynamically generated resonance
arising from meson–baryon interactions. In chiral unitary approaches
based on coupled-channel dynamics, the $\Lambda(1405)$ emerges
naturally through the interaction of the $\bar{K}N$ and $\pi\Sigma$
channels and is frequently interpreted as a quasi-bound $\bar{K}N$
molecular state embedded in the $\pi\Sigma$ continuum.
One of the most remarkable developments associated with the
$\Lambda(1405)$ is the prediction of a two-pole structure in the
complex energy plane. In this picture, the experimentally observed
resonance does not correspond to a single Breit-Wigner state but
instead arises from the superposition of two nearby poles with
different couplings to the $\bar{K}N$ and $\pi\Sigma$ channels. This
interpretation provides a natural explanation for the
channel-dependent line shapes observed in different production
reactions and decay modes. There is recent experimental evidence for the
two-pole structure, and the $\Lambda(1405)$ is now listed as two
states: $\Lambda(1380)\,\frac{1}{2}^-$ and, with the traditional name,
$\Lambda(1405)\,\frac{1}{2}^-$. For more information, see Ref.~[\cite{ParticleDataGroup:2024cfk}] and
the review on the {\it Pole Structure of the $\Lambda(1405)$ Region} therein.

\paragraph{Multi-strange $\Xi$ and $\Sigma$ resonances}
According to the 2026 edition of the
RPP~[\cite{ParticleDataGroup:2024cfk}], the experimentally established
$\Xi$~spectrum contains 12 doubly strange $\Xi$~hyperons, five of which are not
included in the Summary Table, as well as five triply strange
$\Omega^-$~hyperons, of which two are omitted from the Summary
Table. In addition to the recently promoted $\Xi(1820)$~state, only the octet ground states $\Xi^0$ and
$\Xi^-$ and the decuplet states $\Xi(1530)$ and $\Omega^-$ have
achieved a four-star status, indicating solid experimental
evidence. The remaining states listed in the Summary Table carry
three-star ratings, corresponding to evidence ranging from likely to
very likely according to the PDG classification
scheme. Experimental information on decay properties is also extremely
limited, and consequently, apart from the well-established $\Xi^0$, $\Xi^-$, and
$\Omega^-$~states, decay branching fractions have generally not been
measured, and most entries are reported only as upper limits or simply
marked as {\it seen}. The situation is similarly incomplete for spin-parity assignments. Of
the six $\Xi$ resonances with at least a three-star rating, only
$\Xi(1530)\,\frac{3}{2}^+$ and $\Xi(1820)\,\frac{3}{2}^-$ possess
experimental evidence supporting their $J^P$ quantum numbers. For all other $\Xi$ states,
including the $\Xi(1320)$ ground state, the quoted quantum numbers are
inferred primarily from quark-model expectations rather than direct
measurements. In the search for excited states, the 1600--1700~MeV
mass region is of particular importance, especially for resolving the
long-standing $\ast\ast\,\Xi(1620)\,/\ast\ast\ast\,\Xi(1690)$~puzzle. The Belle Collaboration has
provided strong evidence for the existence of the previously poorly
established $\Xi(1620)$~resonance; however, its low mass does
not fit naturally into the conventional baryon spectrum. An important
open question is whether the $\Xi(1620)$ corresponds to a single
resonance. Its comparatively large width may suggest that the observed
structure consists of multiple overlapping states or, alternatively,
that it possesses a more exotic underlying nature.

In the $\Omega$~sector, the quantum numbers of the
$\Omega(1670)$ ground state follow naturally from its placement within
the SU(3) baryon decuplet. However, some experimental support for this assignment
was provided by the BaBar Collaboration, which reported that the
$\Omega^-$~spin is consistent with
$J=\frac{3}{2}$~[\cite{BaBar:2006omx}], assuming spin-$\frac{1}{2}$
assignments for the $\Xi_c^0$ and $\Omega_c^0$. By contrast, the spin
assignments of the remaining excited $\Omega$~resonances remain
largely speculative. The most recent addition to the $\Omega$~family,
the $\Omega(2012)$, was reported by the Belle Collaboration in
2018~[\cite{Belle:2018mqs}]. There is strong evidence that this
resonance carries spin-parity quantum numbers $J^P = \frac{3}{2}^-$;
however, its underlying nature remains an open
question. Interpretations range from a molecular meson-baryon state to
an assignment within the $^2{\bf 10}_{J^-}$~doublet of the
$\big(\,{\bf 70},\,1_1^-\,\big)$~supermultiplet with $J=3/2$. The
possible observation of the corresponding $J=1/2$~partner at a slightly
lower mass could
provide important insight into the structure and classification of
this state.

In recent years, collider experiments have made important
contributions to the study of light doubly and triply strange
hyperons. A variety of excited $\Xi$ states has been observed in
decays of charmed baryons at high-luminosity $e^+e^-$ facilities. The
Belle Collaboration studied the $\Xi(1690)$ in the decay channels
$\Lambda_c^+\to (\Sigma^+K^-)_{\,\Xi(1690)}\,K^+$ and
$\Lambda_c^+\to (\Lambda\bar{K}^{\,0})_{\,\Xi(1690)}\,K^+$~[\cite{Belle:2001hyr}],
while the BaBar Collaboration investigated excited $\Xi^\ast$
resonances in $\Lambda_c^+\to (\Xi^-\pi^+)_{\,\Xi^\ast}\,K^+$
decays~[\cite{BaBar:2008myc}]. Subsequently, Belle significantly
improved the statistical precision of such studies through
measurements of $\Xi_c^+\to (\Xi^-\pi^+)_{\,\Xi^\ast}\,\pi^+$~decays,
providing further evidence for the existence of both $\Xi(1620)$ and $\Xi(1690)$~[\cite{Belle:2018lws}].
Collider experiments have also advanced the spectroscopy of triply
strange hyperons. In particular, the $\Omega(2012)^-$~resonance was
discovered by the Belle Collaboration in the decay modes
$\Omega(2012)^-\to \Xi^0K^-$ and $\Xi^- K_S^0$, produced in decays of
the bottomonium states $\Upsilon(1S)$, $\Upsilon(2S)$, and
$\Upsilon(3S)$. In addition, large samples of ground-state $\Omega^-$
hyperons are produced in weak decays such as $\Xi_c^0\to\Omega^- K^+$,
with smaller contributions from $\Omega_c^0\to\Omega^-\pi^+$. These
data sets have enabled detailed studies of $\Omega^-$ properties,
including the spin measurement performed by the BaBar
Collaboration~[\cite{BaBar:2006omx}]. We refer to Ref.~[\cite{Crede:2024hur}] for a
comprehensive review of multi-strange $\Xi$ and $\Omega$~baryons. 

\subsubsection{Heavy-flavor, charmed and bottom baryons}
In the 2025 update of the Review of Particle Physics, the
Particle Data Group has listed 28 singly charmed baryons, one doubly
charmed baryon, and 27 bottom baryons in the Baryon Summary
Table~[\cite{ParticleDataGroup:2024cfk}] with at least a three-star status. Despite the rapidly growing observed
heavy-baryon spectrum, only three states -- the $\Lambda_c^+$,
$\Sigma_c(2455)$, and $\Xi_c^0$ -- currently possess a four-star
status, indicating there is firm experimental evidence. The development of
bottom-baryon spectroscopy has been particularly striking over the
past decade: the number of listed bottom baryons has increased from
only six states in the 2012 edition of the RPP to 27 in the 2025
compilation. Compared to the 2022 edition alone, eight additional bottom
baryons have been added to the Summary Table, underscoring the rapid
pace of experimental progress in this sector.
Nevertheless, detailed information on many heavy baryons remains
limited. With the exception of the $\Lambda_b^0$, the
$\Big(\,\Xi_b^-,\Xi_b^0\,\Big)$ doublet, and the $\Omega_b^-$, most bottom baryons
have been observed in only a single decay channel, and their decay
status is frequently listed merely as {\it seen}. In many cases, the
isospin, spin, and parity quantum numbers have not yet been determined
experimentally and are instead assigned on the basis of constituent
quark-model expectations. Experimental information on the spin
structure of heavy baryons has only recently begun to improve. For
example, the BES\,III Collaboration measured the spin of the
$\Lambda_c^+$ ground state to be consistent with
$J=\frac{1}{2}$~[\cite{BESIII:2020kap}], while the analysis of 
angular distributions in
several other decay processes has provided evidence that is suggestive of the
spins of higher-mass excited states.

A major breakthrough in heavy-baryon spectroscopy was achieved by the
LHCb Collaboration with the discovery of the doubly charmed baryon
$\Xi_{cc}^{++}$ in 2017~[\cite{LHCb:2017iph}]. The state was observed in
the decay channel $\Xi_{cc}^{++}\to \Lambda_c^+ K^- \pi^+ \pi^+$ with
a measured mass near $3621$~MeV. This observation provided the
first unambiguous evidence for a baryon containing two heavy charm
quarks and one light quark, and opened a new exciting field in hadron
spectroscopy. Doubly heavy baryons are of particular theoretical
interest because they offer a unique laboratory for studying the
interplay between heavy-quark dynamics and light-quark degrees of
freedom in quantum chromodynamics, complementary to the study of
singly heavy baryons. Subsequent measurements by the LHCb
Collaboration have improved the precision of the mass and lifetime
determinations and established additional decay modes of the
$\Xi_{cc}^{++}$. Clearly in the next few years, we can expect more
discoveries in this sector. There are many predictions of the
properties of excited $\Xi_{cc}$~states and other doubly heavy baryons
that will be tested.

For the ground-state singly heavy baryons, the classification follows
naturally from the coupling of the heavy-quark spin $s_Q=\frac{1}{2}$
with the total angular momentum of the light diquark system, $j_{qq}$,
as discussed in Section~\ref{BaryonClassification:HeavyFlavors}. When
the light diquark is in a $j_{qq} = 0$~configuration, there is only a
single ${\bf{\bar{3}}}_F$ flavor multiplet with quantum numbers
$J^P=\frac{1}{2}^+$. In contrast, a light diquark with $j_{qq}=1$
gives rise to two symmetric ${\bf 6}_F$ flavor multiplets with
$J^P=\frac{1}{2}^+$ and $J^P=\frac{3}{2}^+$ as a result of coupling
angular momenta, $J = s_Q \otimes j_{qq}$. Experimentally, nearly all
$S$-wave charmed and bottom ground-state baryons predicted within this
scheme have now been observed, with the notable exceptions of the
$\Xi_b^{\,\prime\,0}\,\frac{1}{2}^+$ and $\Omega_b^\ast\,\frac{3}{2}^+$
states (see Table~\ref{chap3:tab2}).
The situation becomes considerably less clear for orbitally excited
heavy baryons. Assignments of experimentally observed resonances to
specific excitation multiplets remain tentative in many cases, and
numerous predicted states have yet to be identified. Beyond their
masses, production characteristics and decay patterns provide crucial
information for clarifying the internal structure of these systems. In
the charmed baryon sector, the quartet of states $\Lambda_c(2595)$,
$\Lambda_c(2625)$, $\Xi_c(2790)$, and $\Xi_c(2815)$ is widely
interpreted as candidates for the $P$-wave ($l_\lambda=1$)
${\bf{\bar{3}}}_F$ multiplets with $J^P=\frac{1}{2}^-$ and
$\frac{3}{2}^-$. Similarly, the quartet $\Lambda_c(2860)$,
$\Lambda_c(2880)$, $\Xi_c(3055)$, and $\Xi_c(3080)$ provides excellent
candidates for $D$-wave ($l_\lambda=2$) excitations carrying
$J^P=\frac{3}{2}^+$ and $\frac{5}{2}^+$ quantum numbers, thereby completing another
pair of ${\bf{\bar{3}}}_F$ flavor multiplets (see Table~\ref{chap3:tab2}). Nevertheless, several
aspects of these assignments remain under debate.

Our experimental knowledge of the excited $\Sigma_c$ and $\Omega_c$
spectra in the ${\bf 6}_F$ flavor multiplets is still particularly
limited, since the quantum numbers of most observed states have not
yet been determined directly. A similar situation persists in the
singly bottom sector. The pairs $\Lambda_b(5912)$ and
$\Lambda_b(5920)$, as well as $\Lambda_b(6146)$ and $\Lambda_b(6152)$,
are generally interpreted as candidates for the $P$-wave and $D$-wave
${\bf{\bar{3}}}_F$ multiplets with quantum numbers
$J^P=\frac{1}{2}^-$, $\frac{3}{2}^-$ and $J^P=\frac{3}{2}^+$,
$\frac{5}{2}^+$, respectively (see Table~\ref{chap3:tab2}). For most other excited bottom baryons,
however, the spin and parity assignments remain largely unknown, see
the review in Ref.~[\cite{Crede:2024hur}].

\begin{figure}[t]
\centerline{\includegraphics[width=0.98\textwidth]{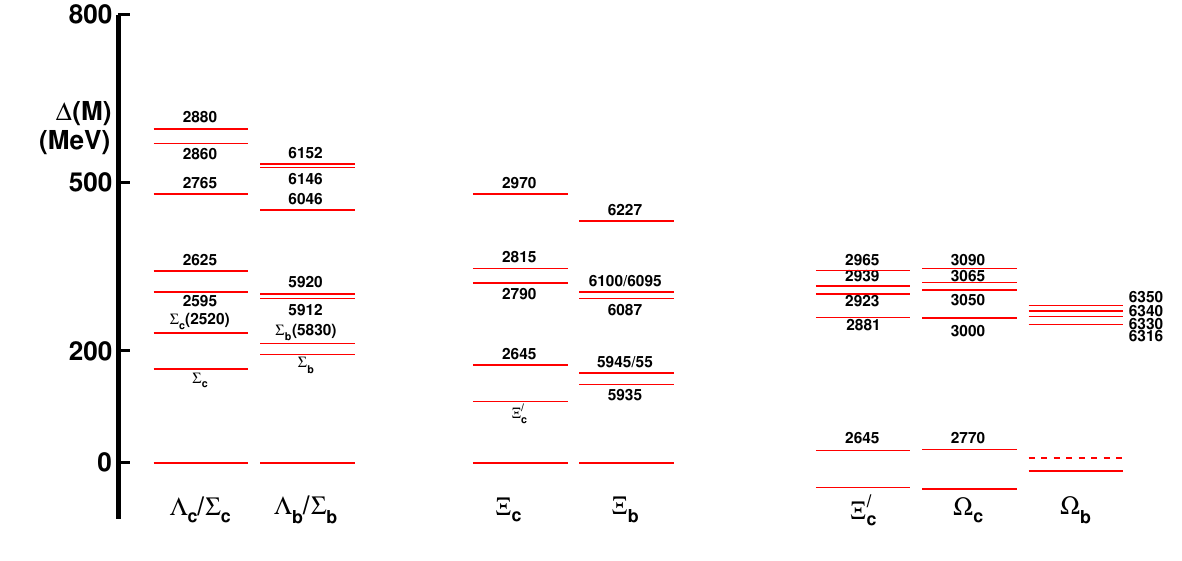}}
\caption{\label{chap3:fig3}Comparison of the (incomplete) charmed baryon and bottom baryon
  spectra. Note that isospin splitting is ignored, and the baseline is given as the weakly decaying
  ground state for the $\Lambda$ and $\Xi$, and the spin-weighted average of the lowest two states
  for the $\Xi_c^{\prime}$ and $\Omega$. The dashed line estimates the position of the first excited
  state of the $\Omega_b$. Reproduced from [\cite{Crede:2024hur}]. \copyright~IOP Publishing Ltd.
  All rights reserved.}
\end{figure}

Figure~\ref{chap3:fig3} illustrates the comparison between the charmed
and bottom baryon spectra. In particular, the expected reduction of
hyperfine splittings with increasing heavy-quark mass, arising from
the approximate inverse dependence on the heavy-quark mass, is clearly
observed across a wide range of states. The spectra also exhibit the
anticipated modest decrease in the excitation energies associated with
$\lambda$-mode excitations when moving from the charm to the bottom
sector. Remarkably, these regularities appear to persist even for
resonances whose spin-parity quantum numbers have not yet been
established experimentally. Consequently, reliable state assignments
often require a simultaneous comparison of both the charm and bottom
spectra rather than an isolated interpretation within a single sector.
As demonstrated in Fig.~\ref{chap3:fig3}, these simple phenomenological
patterns would have allowed the masses of at least 16 excited bottom baryons to be
estimated with considerable accuracy through extrapolations from the
corresponding charmed baryons. Owing to the substantially larger
bottom-quark mass and the correspondingly smaller production cross
sections, the experimental exploration of the bottom baryon spectrum
has progressed more slowly than in the charm sector. It therefore
remains an open question whether counterparts of certain charmed
resonances will eventually be identified in the bottom spectrum.

In conclusion, the role of light-diquark correlations remains an
important question in heavy-baryon spectroscopy, in particular the interplay
between $j_{qq} = 0$ (so-called ``good'') and $j_{qq} = 1$ (so-called ``bad'') diquark
configurations. Such correlations are believed to play a central role
in shaping the singly heavy baryon spectrum, and therefore, they represent a
key topic for future theoretical and experimental
studies. Interestingly, the excited $\Omega_c^\ast$~spectrum may
provide one of the cleanest laboratories for investigating complete
heavy-baryon multiplets. This is somewhat ironic given that the
$\Omega_c$ itself was historically difficult to discover and that
production cross sections for excited $\Omega_c$~states in
$e^+e^-$~collisions are comparatively small. However, once produced, these
resonances are expected to be relatively narrow because the available
decay channels are strongly constrained by isospin conservation. By
contrast, excited $\Lambda_c$~states can decay through intermediate
$\Sigma_c$ and $\Sigma_c^\ast$~channels, leading to broader widths and
more complicated experimental signatures.

\subsubsection{A short story of pentaquarks}
The history of pentaquarks is one of the most fascinating, but also
controversial, chapters in hadron spectroscopy. QCD naturally allows
for ``exotic'' baryons, i.e., baryon-like states
whose valence quark configurations extend beyond the conventional
three-quark picture. The simplest such exotic baryon configuration
is a five-quark color-singlet state, as discussed previously in this
chapter of the Encyclopedia of Nuclear Physics.
The pentaquark story began in earnest in the late 1990s and
early 2000s, when models based on chiral soliton
approaches predicted the existence of a narrow exotic baryon, called the
$\Theta^+(1540)$, with minimal quark content $uudd\bar{s}$. This state
was particularly remarkable because it carried positive strangeness, $S=+1$,
making it manifestly exotic and impossible to describe as an ordinary
three-quark baryon. In 2003, several experimental groups reported
evidence for a narrow resonance near $1540$~MeV. The announcement generated
enormous excitement throughout the hadron physics community and
triggered a worldwide experimental effort. For a short period, the
pentaquark appeared to be one of the most important discoveries in
baryon spectroscopy.
However, the situation soon became controversial. As higher-statistics
measurements became available, many experiments failed to reproduce
the original signal. Dedicated searches at major facilities found no
convincing evidence for the $\Theta^+$, and the claimed evidence was
interpreted as a fluctuation. The resonance
gradually disappeared from the market and is now considered highly
unlikely, and “obituaries” have already appeared in review articles.
The episode
became an important lesson in hadron spectroscopy, highlighting the
challenges associated with low-statistics signals, background
fluctuations, and the interpretation of complex invariant-mass
distributions.

The topic remains intriguing, though. Interest in pentaquarks was revived spectacularly in 2015 when the
LHCb Collaboration reported the observation of two resonant structures
in the decay $\Lambda_b^0 \to J/\psi p K^-$.
The structures, later denoted $P_c(4380)^+$ and $P_c(4450)^+$,
appeared in the invariant $p J/\psi$ mass distribution and exhibited
properties and behaviors consistent with resonances.
Unlike the earlier $\Theta^+$~claims, the statistical significance and
experimental quality of the LHCb data left little doubt that these
states were real. Subsequent analyses with even larger data sets
resolved the original broad structures into several narrower
resonances, including the $P_c(4312)^+$, $P_c(4440)^+$, and
$P_c(4457)^+$ states, which are listed as one-star
$P_{c\bar{c}}$~states in the
RPP~[\cite{ParticleDataGroup:2024cfk}]. Furthermore,
two one-star strange counterparts are also listed, the $P_{c\bar{c}s}(4338)^0\,\frac{1}{2}^-$ and
$P_{c\bar{c}s}(4459)^0$, also reported by the LHCb Collaboration. 

The interpretation of the $P_{c\bar{c}}$~resonances remains an active
and exciting area of research. Some models describe them as compact
five-quark states, while others interpret them as hadronic molecules
formed from weakly bound charmed mesons and baryons, analogous to
nuclear bound states. In particular, the proximity of the
$P_{c\bar{c}}$~masses to the $\Sigma_c^+\bar{D}^{\,0}$ and
$\Sigma_c^+\bar{D}^{\ast\,0}$~thresholds supports the idea of a molecular
interpretation. Regardless of their precise internal structure, the
discovery of the $P_c$ states established beyond doubt that QCD
supports exotic hadrons beyond the traditional quark-model picture.
Today, pentaquark spectroscopy has become a major field within hadron
physics. Ongoing experimental programs at facilities such as the Large
Hadron Collider, Jefferson Laboratory, the future electron-ion
facilities, and future experimental programs at GSI using hadron beams
aim to clarify the spectroscopy, production mechanisms, and
internal dynamics of exotic hadrons. The study of pentaquarks now
provides a powerful new window into the non-perturbative regime of QCD
and the rich spectrum of strongly interacting matter.

\section{An Epilogue to this Story on Hadron Spectroscopy}
Despite decades of significant theoretical and experimental progress, a fully
coherent understanding of the baryon spectrum and baryon structure has not
yet emerged. Different theoretical approaches successfully describe
different aspects of the baryon spectrum. For example, Regge-like trajectories,
characterized by the approximate relation $M^2\propto L+N_{\rm
  \,radial}$, are naturally reproduced in holographic approaches such
as AdS/QCD, see for example Ref.~[\cite{Brodsky:2008pg}]. In contrast, unitary effective field theories provide a
consistent framework for meson-baryon interactions and suggest that
some resonances may arise dynamically from coupled-channel effects
rather than from simple three-quark configurations. At the same time,
constituent quark models remain essential for interpreting the overall
baryon spectrum, including sequential decays through intermediate
resonant states, and the
long-standing problem of missing resonances.
The intrinsic symmetry of quark pairs also appears to play an
important role in determining baryon masses. In particular, whether
two quarks combine symmetrically or antisymmetrically under exchange
can significantly influence the resulting spectrum. While such effects
may be interpreted in terms of effective gluon exchange, they are
often more naturally associated with non-perturbative QCD dynamics,
including quark and gluon condensates and instanton-induced
interactions.

Recent high-precision photoproduction experiments, especially those
involving polarized beams and targets, have led to the discovery of
new baryon resonances and substantially improved our knowledge of
previously known states. Nevertheless, the overall picture remains
incomplete, demonstrating the intrinsic complexity of quantum
chromodynamics in the non-perturbative regime. Rapid advances are
expected from both experiment and theory using advanced computing
technologies and AI-driven techniques. In particular, increasingly
sophisticated lattice-QCD calculations are anticipated to provide
valuable insight into the structure and dynamics of baryon
resonances. On the experimental side, further measurements of
polarization observables in photo- and electroproduction processes,
including reactions on neutron targets, will be crucial. Additional
information is also expected from studies of multi-meson final states,
strange baryon spectroscopy, $p\bar{p}$~annihilation, reactions
induced by pion and Kaon beams,  and baryon production in $J/\psi$ and
$\psi^{\,\prime}$~decays. Together, these efforts will help deepen our
understanding of baryons as bound states governed by the strong interaction.

Hadron spectroscopy remains an exciting, active, and lively field in
nuclear and high-energy (particle) physics.
But despite the tremendous progress achieved over recent decades, hadron
spectroscopy also continues to pose fundamental questions about the
dynamics of confinement and the emergence of hadronic structure from
QCD. The steadily growing body of high-precision data, together with
advances in partial-wave analyses, lattice-QCD calculations, and
effective field-theory approaches, has significantly improved our
understanding of conventional and exotic hadrons alike. At the same
time, many challenges remain, including the identification of missing
meson and baryon
resonances, the interpretation of multiquark candidates, the search
for exotic hadrons, and the
determination of the relevant effective degrees of freedom governing
hadron structure. Future experiments at high-intensity hadron and
electron facilities, combined with increasingly sophisticated
theoretical methods, are expected to provide deeper insight into the
non-perturbative regime of QCD and the rich excitation spectrum of
strongly interacting matter.

\begin{ack}[Acknowledgments]

This work was partially supported by the Department of Energy, Office of
Science, Office of Nuclear Physics under Contract No. DE-FG02-92ER40735.
\end{ack}


\bibliographystyle{Harvard}
\bibliography{HadronSpectroscopy}

@article{D0:1995jca,
    author = "Abachi, S. and others",
    collaboration = "D0",
    title = "{Observation of the top quark}",
    archivePrefix = "arXiv",
    reportNumber = "FERMILAB-PUB-95-028-E, D0-2698, D0-PUB-95-7",
    doi = "10.1103/PhysRevLett.74.2632",
    journal = "Phys. Rev. Lett.",
    volume = "74",
    pages = "2632--2637",
    year = "1995"
}

@article{LHCb:2015yax,
    author = "Aaij, Roel and others",
    collaboration = "LHCb",
    title = "{Observation of $J/\psi p$ Resonances Consistent with Pentaquark States in $\Lambda_b^0 \to J/\psi K^- p$ Decays}",
    archivePrefix = "arXiv",
    primaryClass = "hep-ex",
    reportNumber = "CERN-PH-EP-2015-153, LHCB-PAPER-2015-029",
    doi = "10.1103/PhysRevLett.115.072001",
    journal = "Phys. Rev. Lett.",
    volume = "115",
    pages = "072001",
    year = "2015"
}

@article{COMPASS:2014vkj,
    author = "Adolph, C. and others",
    collaboration = "COMPASS",
    title = "{Odd and even partial waves of $\eta\pi^-$ and $\eta'\pi^-$ in $\pi^-p\to\eta^{(\prime)}\pi^-p$ at $191\,\textrm{GeV}/c$}",
    archivePrefix = "arXiv",
    primaryClass = "hep-ex",
    reportNumber = "CERN-PH-EP-2014-204",
    doi = "10.1016/j.physletb.2014.11.058",
    journal = "Phys. Lett. B",
    volume = "740",
    pages = "303--311",
    year = "2015",
    note = "[Erratum: Phys.Lett.B 811, 135913 (2020)]"
}

@article{JPAC:2018zyd,
    author = "Rodas, A. and others",
    collaboration = "JPAC",
    title = "{Determination of the pole position of the lightest hybrid meson candidate}",
    archivePrefix = "arXiv",
    primaryClass = "hep-ph",
    reportNumber = "JLAB-THY-18-2839",
    doi = "10.1103/PhysRevLett.122.042002",
    journal = "Phys. Rev. Lett.",
    volume = "122",
    number = "4",
    pages = "042002",
    year = "2019"
}

@article{ParticleDataGroup:2024cfk,
    author = "Navas, S. and others",
    collaboration = "Particle Data Group",
    title = "{Review of particle physics}",
    doi = "10.1103/PhysRevD.110.030001",
    journal = "Phys. Rev. D",
    volume = "110",
    number = "3",
    pages = "030001",
    year = "2024"
}

@article{ParticleDataGroup:1984mfx,
    author = "Wohl, C. G. and others",
    collaboration = "Particle Data Group",
    title = "{Review of Particle Properties. Particle Data Group}",
    doi = "10.1103/RevModPhys.56.S1",
    journal = "Rev. Mod. Phys.",
    volume = "56",
    pages = "S1--S304",
    year = "1984"
}

@article{BESIII:2022riz,
    author = "Ablikim, M. and others",
    collaboration = "BESIII",
    title = "{Observation of an Isoscalar Resonance with Exotic $J^{PC}=1^{-+}$ Quantum Numbers in $J/\psi\to\gamma\eta\eta^{\,\prime}$}",
    archivePrefix = "arXiv",
    primaryClass = "hep-ex",
    doi = "10.1103/PhysRevLett.129.192002",
    journal = "Phys. Rev. Lett.",
    volume = "129",
    number = "19",
    pages = "192002",
    year = "2022",
    note = "[Erratum: Phys.Rev.Lett. 130, 159901 (2023)]"
}

@article{BESIII:2023wfi,
    author = "Ablikim, Medina and others",
    collaboration = "BESIII",
    title = "{Determination of Spin-Parity Quantum Numbers of X(2370) as $0^{-+}$ from $J/\psi\to\gamma K_S^0 K_S^0 \eta^{\,\prime}$}",
    archivePrefix = "arXiv",
    primaryClass = "hep-ex",
    doi = "10.1103/PhysRevLett.132.181901",
    journal = "Phys. Rev. Lett.",
    volume = "132",
    number = "18",
    pages = "181901",
    year = "2024"
}

@article{JPAC:2021rxu,
    author = "Albaladejo, Miguel and others",
    collaboration = "JPAC",
    title = "{Novel approaches in hadron spectroscopy}",
    eprint = "2112.13436",
    archivePrefix = "arXiv",
    primaryClass = "hep-ph",
    reportNumber = "LA-UR-21-31664, JLAB-THY-22-3459",
    doi = "10.1016/j.ppnp.2022.103981",
    journal = "Prog. Part. Nucl. Phys.",
    volume = "127",
    pages = "103981",
    year = "2022"
}

@article{Gross:2022hyw,
    author = "Gross, Franz and others",
    title = "{50 Years of Quantum Chromodynamics}",
    eprint = "2212.11107",
    archivePrefix = "arXiv",
    primaryClass = "hep-ph",
    doi = "10.1140/epjc/s10052-023-11949-2",
    journal = "Eur. Phys. J. C",
    volume = "83",
    pages = "1125",
    year = "2023"
}

@article{Chen:2016spr,
    author = "Chen, Hua-Xing and Chen, Wei and Liu, Xiang and Liu, Yan-Rui and Zhu, Shi-Lin",
    title = "{A review of the open charm and open bottom systems}",
    eprint = "1609.08928",
    archivePrefix = "arXiv",
    primaryClass = "hep-ph",
    doi = "10.1088/1361-6633/aa6420",
    journal = "Rept. Prog. Phys.",
    volume = "80",
    number = "7",
    pages = "076201",
    year = "2017"
}

@article{Jaffe:2004ph,
    author = "Jaffe, R. L.",
    editor = "Kunihiro, Teiji and Onogi, Tetsuya and Abuki, H. and Takahashi, Toru T.",
    title = "{Exotica}",
    eprint = "hep-ph/0409065",
    archivePrefix = "arXiv",
    reportNumber = "MIT-CTP-3538",
    doi = "10.1016/j.physrep.2004.11.005",
    journal = "Phys. Rept.",
    volume = "409",
    pages = "1--45",
    year = "2005"
}

@article{Deur:2018roz,
    author = "Deur, Alexandre and Brodsky, Stanley J. and De T{\'e}ramond, Guy F.",
    title = "{The Spin Structure of the Nucleon}",
    eprint = "1807.05250",
    archivePrefix = "arXiv",
    primaryClass = "hep-ph",
    reportNumber = "JLAB-PHY-18-2760, SLAC-PUB-17279",
    doi = "10.1088/1361-6633/ab0b8f",
    journal = "Rept. Prog. Phys.",
    volume = "82",
    pages = "076201",
    year = "2019"
}

@article{Hillert:2006yb,
    author = "Hillert, W.",
    editor = "Arenhoevel, Hartmuth and Backe, Hartmut and Drechsel, Dieter and Friedrich, Jorg and Kaiser, Karl-Heinz and Walcher, Thomas",
    title = "{The Bonn electron stretcher accelerator ELSA: Past and future}",
    doi = "10.1140/epja/i2006-09-015-4",
    journal = "Eur. Phys. J. A",
    volume = "28S1",
    pages = "139--148",
    year = "2006"
}

@article{Mecking:2006yh,
    author = "Mecking, B. A.",
    editor = "Arenhoevel, Hartmuth and Backe, Hartmut and Drechsel, Dieter and Friedrich, Jorg and Kaiser, Karl-Heinz and Walcher, Thomas",
    title = "{Twenty years of physics at MAMI: What did it mean?}",
    reportNumber = "JLAB-PHY-06-530",
    doi = "10.1140/epja/i2006-09-020-7",
    journal = "Eur. Phys. J. A",
    volume = "28S1",
    pages = "209--219",
    year = "2006"
}

@article{GRAAL:2005mor,
    author = "Bartalini, O. and others",
    collaboration = "GRAAL",
    title = "{Measurement of $\pi^0$ photoproduction on the proton from 550 MeV to 1500 MeV at GRAAL}",
    doi = "10.1140/epja/i2005-10191-2",
    journal = "Eur. Phys. J. A",
    volume = "26",
    pages = "399--419",
    year = "2005"
}

@article{LEPS:2013tbd,
    author = "Muramatsu, N. and others",
    collaboration = "LEPS",
    title = "{Development of high intensity laser-electron photon beams up to 2.9 GeV at the SPring-8 LEPS beamline}",
    eprint = "1308.6453",
    archivePrefix = "arXiv",
    primaryClass = "physics.acc-ph",
    doi = "10.1016/j.nima.2013.11.039",
    journal = "Nucl. Instrum. Meth. A",
    volume = "737",
    pages = "184--194",
    year = "2014"
}

@article{McKeown:2013sxa,
    author = "McKeown, R. D.",
    editor = "Goity, Jose and Chen, Jianping",
    title = "{Overview of Nuclear Physics at Jefferson Lab}",
    eprint = "1303.6622",
    archivePrefix = "arXiv",
    primaryClass = "physics.acc-ph",
    reportNumber = "JLAB-PHY-13-1691, DOE-OR-23177-2487",
    doi = "10.22323/1.172.0001",
    journal = "PoS",
    volume = "CD12",
    pages = "001",
    year = "2013"
}

@article{Chew:1957tf,
    author = "Chew, G. F. and Goldberger, M. L. and Low, F. E. and Nambu, Yoichiro",
    title = "{Relativistic dispersion relation approach to photomeson production}",
    doi = "10.1103/PhysRev.106.1345",
    journal = "Phys. Rev.",
    volume = "106",
    pages = "1345--1355",
    year = "1957"
}

@article{Klempt:2020bdu,
    author = "Klempt, E. and Burkert, V. and Thoma, U. and Tiator, L. and Workman, R.",
    collaboration = "Baryon@PDG Group",
    title = "{$\varLambda $ and $\varSigma $ Excitations and the Quark Model}",
    eprint = "2007.04232",
    archivePrefix = "arXiv",
    primaryClass = "nucl-ex",
    doi = "10.1140/epja/s10050-020-00261-2",
    journal = "Eur. Phys. J. A",
    volume = "56",
    number = "10",
    pages = "261",
    year = "2020"
}

@article{CLAS:2014tbc,
    author = "Moriya, K. and others",
    collaboration = "CLAS",
    title = "{Spin and parity measurement of the $\Lambda(1405)$ baryon}",
    eprint = "1402.2296",
    archivePrefix = "arXiv",
    primaryClass = "hep-ex",
    reportNumber = "JLAB-PHY-14-1848",
    doi = "10.1103/PhysRevLett.112.082004",
    journal = "Phys. Rev. Lett.",
    volume = "112",
    number = "8",
    pages = "082004",
    year = "2014"
}

@article{BaBar:2006omx,
    author = "Aubert, Bernard and others",
    editor = "Sissakian, Alexey and Kozlov, Gennady and Kolganova, Elena",
    collaboration = "BaBar",
    title = "{Measurement of the spin of the $\Omega^-$ hyperon at BABAR}",
    eprint = "hep-ex/0606039",
    archivePrefix = "arXiv",
    reportNumber = "SLAC-PUB-11896, BABAR-PUB-05-05, BABAR-PUB-06-034",
    doi = "10.1103/PhysRevLett.97.112001",
    journal = "Phys. Rev. Lett.",
    volume = "97",
    pages = "112001",
    year = "2006"
}

@article{Belle:2001hyr,
    author = "Abe, Kazuo and others",
    collaboration = "Belle",
    title = "{Observation of Cabibbo suppressed and $W$~exchange $\Lambda^+_c$ baryon decays}",
    eprint = "hep-ex/0111032",
    archivePrefix = "arXiv",
    reportNumber = "BELLE-PREPRINT-2001-17, KEK-PREPRINT-2001-137",
    doi = "10.1016/S0370-2693(01)01373-9",
    journal = "Phys. Lett. B",
    volume = "524",
    pages = "33--43",
    year = "2002"
}

@article{BaBar:2008myc,
    author = "Aubert, Bernard and others",
    collaboration = "BaBar",
    title = "{Measurement of the Spin of the $\Xi(1530)$ Resonance}",
    eprint = "0803.1863",
    archivePrefix = "arXiv",
    primaryClass = "hep-ex",
    reportNumber = "SLAC-PUB-13160, BABAR-PUB-07-073",
    doi = "10.1103/PhysRevD.78.034008",
    journal = "Phys. Rev. D",
    volume = "78",
    pages = "034008",
    year = "2008"
}

@article{Belle:2018lws,
    author = "Sumihama, M. and others",
    collaboration = "Belle",
    title = "{Observation of $\Xi(1620)^0$ and evidence for $\Xi(1690)^0$ in $\Xi_c^+ \rightarrow \Xi^-\pi^+\pi^+$ decays}",
    eprint = "1810.06181",
    archivePrefix = "arXiv",
    primaryClass = "hep-ex",
    doi = "10.1103/PhysRevLett.122.072501",
    journal = "Phys. Rev. Lett.",
    volume = "122",
    number = "7",
    pages = "072501",
    year = "2019"
}

@article{Crede:2024hur,
    author = "Crede, Volker and Yelton, John",
    title = "{70 years of hyperon spectroscopy: a review of strange $\Xi$, $\Omega$ baryons, and the spectrum of charmed and bottom baryons}",
    eprint = "2502.08815",
    archivePrefix = "arXiv",
    primaryClass = "hep-ex",
    doi = "10.1088/1361-6633/ad7610",
    journal = "Rept. Prog. Phys.",
    volume = "87",
    number = "10",
    pages = "106301",
    year = "2024"
}

@article{BESIII:2020kap,
    author = "Ablikim, Medina and others",
    collaboration = "BESIII",
    title = "{Determination of the  $\Lambda_c^+$ spin via the reaction $e^+e^-\to\Lambda_c^+\bar\Lambda_c^-$}",
    eprint = "2011.00396",
    archivePrefix = "arXiv",
    primaryClass = "hep-ex",
    doi = "10.1103/PhysRevD.103.L091101",
    journal = "Phys. Rev. D",
    volume = "103",
    number = "9",
    pages = "L091101",
    year = "2021"
}

@article{LHCb:2017iph,
    author = "Aaij, Roel and others",
    collaboration = "LHCb",
    title = "{Observation of the doubly charmed baryon $\Xi_{cc}^{++}$}",
    eprint = "1707.01621",
    archivePrefix = "arXiv",
    primaryClass = "hep-ex",
    reportNumber = "LHCB-PAPER-2017-018, CERN-EP-2017-156",
    doi = "10.1103/PhysRevLett.119.112001",
    journal = "Phys. Rev. Lett.",
    volume = "119",
    number = "11",
    pages = "112001",
    year = "2017"
}

@article{Dudek:2011bn,
    author = "Dudek, Jozef J.",
    title = "{The lightest hybrid meson supermultiplet in QCD}",
    eprint = "1106.5515",
    archivePrefix = "arXiv",
    primaryClass = "hep-ph",
    reportNumber = "JLAB-THY-11-1387",
    doi = "10.1103/PhysRevD.84.074023",
    journal = "Phys. Rev. D",
    volume = "84",
    pages = "074023",
    year = "2011"
}

@article{Edwards:2012fx,
    author = "Edwards, Robert G. and Mathur, Nilmani and Richards, David G. and Wallace, Stephen J.",
    collaboration = "Hadron Spectrum",
    title = "{Flavor structure of the excited baryon spectra from lattice QCD}",
    eprint = "1212.5236",
    archivePrefix = "arXiv",
    primaryClass = "hep-ph",
    reportNumber = "JLAB-THY-12-1680, TIFR-TH-12-50",
    doi = "10.1103/PhysRevD.87.054506",
    journal = "Phys. Rev. D",
    volume = "87",
    number = "5",
    pages = "054506",
    year = "2013"
}

@article{Glozman:2003bt,
    author = "Glozman, L. Ya.",
    title = "{Chiral multiplets of excited mesons}",
    eprint = "hep-ph/0312354",
    archivePrefix = "arXiv",
    doi = "10.1016/j.physletb.2004.02.066",
    journal = "Phys. Lett. B",
    volume = "587",
    pages = "69--77",
    year = "2004"
}

@article{Kaiser:1995cy,
    author = "Kaiser, Norbert and Siegel, P. B. and Weise, W.",
    title = "{Chiral dynamics and the $S_{11}(1535)$ nucleon resonance}",
    eprint = "nucl-th/9507036",
    archivePrefix = "arXiv",
    reportNumber = "TUM-T39-95-10",
    doi = "10.1016/0370-2693(95)01203-3",
    journal = "Phys. Lett. B",
    volume = "362",
    pages = "23--28",
    year = "1995"
}

@article{Bruns:2010sv,
    author = "Bruns, Peter C. and Mai, Maxim and Meissner, Ulf G.",
    title = "{Chiral dynamics of the $S_{11}(1535)$ and $S_{11}(1650)$ resonances revisited}",
    eprint = "1012.2233",
    archivePrefix = "arXiv",
    primaryClass = "nucl-th",
    reportNumber = "HISKP-TH-10-28, FZJ-IKP-TH-2010-25",
    doi = "10.1016/j.physletb.2011.02.008",
    journal = "Phys. Lett. B",
    volume = "697",
    pages = "254--259",
    year = "2011"
}

@article{Chiang:1996em,
    author = "Chiang, Wen-Tai and Tabakin, Frank",
    title = "{Completeness rules for spin observables in pseudoscalar meson photoproduction}",
    eprint = "nucl-th/9611053",
    archivePrefix = "arXiv",
    doi = "10.1103/PhysRevC.55.2054",
    journal = "Phys. Rev. C",
    volume = "55",
    pages = "2054--2066",
    year = "1997"
}

@article{Belle:2018mqs,
    author = "Yelton, J. and others",
    collaboration = "Belle",
    title = "{Observation of an Excited $\Omega^-$ Baryon}",
    eprint = "1805.09384",
    archivePrefix = "arXiv",
    primaryClass = "hep-ex",
    reportNumber = "BELLE-PREPRINT-2018-09, KEK-PREPRINT-2018-3, Belle Preprint 2018-09, KEK Preprint 2018-3",
    doi = "10.1103/PhysRevLett.121.052003",
    journal = "Phys. Rev. Lett.",
    volume = "121",
    number = "5",
    pages = "052003",
    year = "2018"
}

@article{Gross:1973id,
    author = "Gross, David J. and Wilczek, Frank",
    title = "{Ultraviolet Behavior of Nonabelian Gauge Theories}",
    doi = "10.1103/PhysRevLett.30.1343",
    journal = "Phys. Rev. Lett.",
    volume = "30",
    pages = "1343--1346",
    year = "1973"
}

@article{Brodsky:2008pg,
    author = "Brodsky, Stanley J. and de Teramond, Guy F.",
    editor = "Zichichi, Antonino",
    title = "{AdS/CFT and Light-Front QCD}",
    eprint = "0802.0514",
    archivePrefix = "arXiv",
    primaryClass = "hep-ph",
    reportNumber = "SLAC-BPUB-13107",
    doi = "10.1142/9789814293242_0008",
    journal = "Subnucl. Ser.",
    volume = "45",
    pages = "139--183",
    year = "2009"
}

@article{Crede:2013kia,
    author = "Crede, V. and Roberts, W.",
    title = "{Progress towards understanding baryon resonances}",
    eprint = "1302.7299",
    archivePrefix = "arXiv",
    primaryClass = "nucl-ex",
    doi = "10.1088/0034-4885/76/7/076301",
    journal = "Rept. Prog. Phys.",
    volume = "76",
    pages = "076301",
    year = "2013"
}

\end{document}